\documentclass[useAMS,usenatbib,usegraphicx,referee,a4paper]{mn2e}
\newcommand {\aplt} {{\raise-.5ex\hbox{$\buildrel<\over\sim$}}} 
\usepackage{amsmath}
\usepackage{url}
\DeclareMathOperator{\sgn}{sgn}
 \title[Partially resolved binaries in PS1]{Identification of partially resolved binaries in Pan-STARRS\,1 data}
 \author[N.R.\ Deacon et al.]{N.R.\ Deacon\thanks{E-mail:n.deacon2@herts.ac.uk}$^{1,2}$, E.A.\ Magnier$^3$, William M.J.\ Best$^3$, Michael C.\ Liu$^3$,\newauthor T.J. Dupuy$^4$, K.C. Chambers$^3$, P.W. Draper$^5$, H. Flewelling $^3$,\newauthor N. Metcalfe$^5$, J.L. Tonry$^3$,  R.J. Wainscoat $^3$, C. Waters$^3$\\
 $^1$Centre for Astrophysics Research, University of Hertfordshire, College Lane, Hatfield, AL10 9AB, UK\\
 $^2$Max Planck Institute for Astronomy, Konigstuhl 17, Heidelberg, 69117, Germany\\
 $^3$Institute for Astronomy, University of Hawaii at Manoa, 2680 Woodlawn Drive, Honolulu, HI, 96822, USA\\
 $^4$Department of Astronomy, The University of Texas at Austin, 2515 Speedway, Austin, TX 78712, USA\\
 $^5$Department of Physics, University of Durham, South Road, Durham DH1 3LE, UK\\
}
 \begin{document}
 \date{}
 \pagerange{\pageref{firstpage}--\pageref{lastpage}} \pubyear{2015}
 \maketitle
 \label{firstpage}
 \begin{abstract}Using shape measurement techniques developed for weak lensing surveys we have identified three new ultracool binaries in the Pan-STARRS\,1 survey. Binary companions which are not completely resolved can still alter the shapes of stellar images. These shape distortions can be measured if PSF anisotropy caused by the telescope is properly accounted for. We show using both a sample of known binary stars and simulated binaries that we can reliably recover binaries wider than around 0.3\arcsec\,\, and with flux ratios greater than around 0.1. We then applied our method to a sample of ultracool dwarfs within 30\,pc with 293 objects having sufficient Pan-STARRS\,1 data for our method. In total we recovered all but one of the 11 binaries wider than 0.3\arcsec\,\, in this sample. Our one failure was a true binary detected with a significant but erroneously high ellipticity which led it to be rejected in our analysis. We identify three new binaries, one a simultaneous discovery, with primary spectral types M6.5, L1 and T0.5. These latter two were confirmed with Keck/NIRC2 follow-up imaging. This technique will be useful for identifying large numbers of stellar and substellar binaries in the upcoming LSST and DES sky surveys.
 \end{abstract}
 \begin{keywords} \end{keywords}
\section{Introduction}
Double stars, both coincident alignments and true physical systems, are common in the sky. These objects often present opportunities e.g. with binary systems serving as excellent benchmarks to characterise substellar evolutionary models \citep{Liu2008}. Binary stars have a common age and thus can be used to test the accuracy of stellar and substellar evolution models. The statistical properties of large samples of binary stars also represent a key output metric for star formation models. There are also problems introduced by binarity such as the effect that hidden secondary components have on determinations of the initial mass function (see \citealt{Chabrier2003a} for an example of how significant this can be). Transit surveys for exoplanets can be contaminated by stellar blends \citep{Sirko2003} as a background eclipsing binary blended with the target star can induce an erroneous planetary transit detection. Similarly the radius of a real planet orbiting one component of an unidentified stellar binary may be significantly underestimated. Hence large campaigns of both seeing-limited and adaptive-optics observations have been undertaken to weed out stellar blends from samples of candidate exoplanet host stars (see for example \citealt{Law2014}).

A novel method for detecting stellar binaries proposed by \cite{Hoekstra2005} uses image shape analysis developed for weak lensing detections in extragalactic astrophysics \citep{Kaiser1994,Hoekstra1998} to identify stars with pronounced ellipticity, implying two sources blended together. \cite{Terziev2013} expanded this method to wide-field multi-epoch surveys, specifically the Palomar Transient Factory (PTF; \citealt{Law2009}). They demonstrated that both ellipticity and the trend of increasing ellipticity with better seeing could identify candidate double stars. These were confirmed using Robo-AO \citep{Baranec2012} follow-up observations, demonstrating that the vast majority of stellar images identified as elliptical in their test sample were stellar blends and that the majority of stellar images with negligible ellipticity had no resolvable companion.

In this paper we present an initial application of this method to Pan-STARRS\,1 data \citep{Chambers2016}, showing that the stellar binaries identified by \cite{Terziev2013} can be readily identified using Pan-STARRS\,1. We then apply this technique to a sample of bright, nearby L and T dwarfs and identify three new binaries, one each of spectral types M, L and T. Finally we discuss how this technique can be used with the Pan-STARRS\,1 database as a whole, allowing users to identify partially resolved binaries for any sample of input objects. This method will allow large samples of stars to be screened for binarity in the 0.3--1.5\arcsec\, separation range, enabling cleaner samples for exoplanet transit studies such as {\it Kepler K2} \citep{Howell2014}. This is a larger lower resolution bound than space-based telescopes such as Gaia (20\,milliarcseconds \url{http://sci.esa.int/gaia/31441-binary-stars/}) and the $\sim$15\,milliarcseconds possible from advanced aperture masking techniques used in ground-based observations \cite{Kraus2012a}. The planned Large Synoptic Survey Telescope (LSST) will provide sharper, more frequent sampling of the sky allowing this technique to be pushed to even smaller separations.

In Section~2 we use the test sample from \cite{Terziev2013} and simulated images to test to potential of this technique with Pan-STARRS\,1 data. In Section~3 we use this technique to identify new binaries in a sample of nearby ultracool dwarfs. In Section~4 we examine how this technique can be incorporated into the Pan-STARRS\,1 database.

\section{Shape Measurement}
Shape measurement of astronomical objects has long been used to determine object morphology and multiplicity. The recent boom in cosmological parameter estimation based on weak gravitational lensing is built on the shape measurement formalisms developed by \cite{Kaiser1994} and \cite{Hoekstra1998}. These imagine an idealised astronomical image above the atmosphere being distorted by both the atmosphere and telescope and decompose this distortion into two components, shear and smear. Shear is the stretching of an image by gravitational lensing or telescope while smear is the fattening of the image caused by either seeing in the atmosphere or by the telescope optics. While we expect no significant gravitational shear on stars in the solar neighbourhood, the anisotropy introduced by the telescope optics will lead to point-like objects having significant ellipticities. \cite{Hoekstra2005} demonstrated that the weak lensing formalism can be used to correct for the shearing effect of the atmosphere and telescope to measure corrected ellipticities of stellar images. 

Ellipicities are dimensionless numbers produced by combinations of the second position moments of the flux. These moments are defined thus,
\begin{equation}
\label{I_def}
\begin{tabular}{ccc}
$I_{11}=\sum f(x,y) x^2 W(x,y)$&$I_{12}=\sum f(x,y) xy W(x,y)$&$I_{22}=\sum f(x,y) y^2 W(x,y)$\\
\end{tabular}
\end{equation}

Here x and y are pixel positions with respect to the star's photocentre, f(x,y) is the flux in a particular pixel and W(x,y) is a weighting function. This latter term suppresses values further from the photocentre to prevent noisy, low signal to noise data dominating the moments. These moments are combined to form two dimensionless ellipticity parameters,

\begin{equation}
\label{e_def}
\begin{tabular}{cc}
$e_1=\frac{I_{11}-I_{22}}{I_{11}+I_{22}}$&$e_2=\frac{I_{12}}{I_{11}+I_{22}}$\\
\end{tabular}
\end{equation}

The $e_1$ ellipticity polarisation represents elongations along the R.A. and Declination axes, i.e. the "$+$" polarisation; while the $e_2$ ellipticity polarisation represents elongations on axes tilted by 45$^{\circ}$ , the $\times$ polarisation. These ellipticity values can be positive or negative with more elliptical stars having higher total ellipticities ($e_{tot}$, the quadrature sum of $e_1$ and $e_2$). The individual ellipticity parameters are themselves a function of total ellipiticty and the position angle of the binary. If we consider a binary with position angle $\theta$ measured as $\theta=\tan^{-1}\frac{x}{y}$ then the individual ellipticity values till take the form,

\begin{equation}
\label{e_rot}
\begin{tabular}{cc}
$e_1=-e_{tot}\cos{2 \theta}$&$e_2=e_{tot}\sin{2 \theta}$\\
\end{tabular}
\end{equation}

The values of $e_1$ and $e_2$ are symmetric with a 180$^{\circ}$ rotation. See \cite{Terziev2013}'s Section 5.2 for a similar derivation of the relationship between ellipticity values and position angle. Note in later sections we will see a subtly different version of this formula as the result of mapping R.A. to Pan-STARRS\,1 image pixel number. The remainder of this section briefly describes the corrections made to the values of $e_1$ and $e_2$ by \cite{Hoekstra2005}. Any reader interested in the technical details of this method should refer to that paper. Note below we use the suffix $\nu$ to refer to both the ellipticity polarisations. Thus $e_{\nu}$ is a vector with two components, $e_1$ and $e_2$, $p_{\nu}$ contains $p_1$ and $p_2$ etc.

The telescope and the atmosphere will affect a point source image in two ways. Firstly, there is the general smearing of the image caused by seeing or telescope optics (the smear term). Then there is the anisotropy caused largely by the telescope's non-axis symetric PSF (the shear term). The correction for these two effects for point sources is, somewhat counterintuitively, well approximated by subtracting the product of the smear polarisability $P_{sm,\nu\nu}$ for the target (this is derived from various moments of the image of the source) and $p_{\nu}$, a measure of PSF anisotropy at the position of the source on the image. To measure this we measure $p_{\nu}$ for stars across the image and then determine how this parameter changes across the image (which we will call $p_{\nu, smooth}$). We use two different methods, one in our tests on specific targets  (see Section~\ref{im_test}) and one in our application to the general Pan-STARRS\,1 dataset (see Section~\ref{db_test}). Once $p_{\nu}$ has been determined we can measure the polarisation of each object by

\begin{equation}
\label{e_cor}
e_{\nu, cor}= e_{\nu} - P_{sm,\nu\nu}p_{\nu,smooth}
\end{equation}

Where $p_{\nu,smooth}$ is some smoothed function of the $p_{\nu}$ values across the image.

However this is not sufficient to correctly determine the anisotropy if the object is two blended point sources, as there may be significant higher-order moments. Here an additional term  called $\alpha$ is required. This was introduced by \cite{Hoekstra2005} and is an additional correction required for an object which potentially consists of two blended point sources. This modifies Equation~\ref{e_cor} to become,
\begin{equation}
\label{p_smooth}
e_{\nu, cor}= e_{\nu} - P_{sm,\nu\nu}p_{\nu,smooth} -\alpha \sqrt{p_1^2+p_2^2}
\end{equation}

This latter $\alpha$ term is a function of separation and flux ratio. Since these cannot be independently determined for a blended binary with no additional high-resolution imaging, $\alpha$ must be fitted as a free parameter for each object (note that it is a {\it per object} parameter, not a {\it per image} parameter). We will return to how to fit this parameter later.

For a true marginally blended binary, the ellipticity should be a function of seeing. Hence the various values of $e_{\nu, cor}$ are used to constrain a fit of $e_{\nu}$ as a function of seeing. The ellipticity is then measured at a reference seeing value for comparsion with other objects.

The mathematical techniques for measuring the shapes of objects are comprehensively laid out in Appendix~A of see \citealt{Hoekstra1998}.  \cite{Terziev2013} repeats the relevant terms for measuring stellar shapes (albeit with an error in their Equation 6  -- see equation 20 of \cite{Hoekstra1998} for the correct term).

\section{Testing of Pan-STARRS\,1 images}
\label{im_test}
We first applied this technique to Pan-STARRS\,1 images using the same test sample as \cite{Terziev2013}. This sample has been followed up with AO imaging, meaning any of these objects which are binaries with separations down to a few tenths of an arcsecond will have been identified. Hence we can use this sample to measure how well our method detects real astronomical binaries and how often single stars are falsely identified as binaries.

We began by extracting 10'$\times$10' images for each of the 44 objects in the Terziev sample from the Pan-STARRS\,1 postage stamp server. We requested all single-epoch ''warp" images (single epoch images re-registered to a fixed R.A. and Dec. grid with 0.25\arcsec\, pixels) for all filters (the standard $g_{P1}$,$r_{P1}$,$i_{P1}$,$z_{P1}$,$y_{P1}$ filters plus the wider $w_{P1}$ filter used for asteroid searches; \citealt{Tonry2012}) and also extracted the pixel masks for each warp. We then ran the SExtractor software package \citep{SExtractor} on the image to identify sources in each image and to measure their flux, R.A. and Dec., and CCD $X$ and $Y$ positions along with the flags measured by SExtractor. Note that we effectively turned off SExtractor deblending by setting the deblending contrast ratio DEBLEND\_MINCONT to 0.5, to avoid the centroids reported by SExtractor for blended objects from jumping around due to variations in seeing (being resolved in some images and not in others). The Pan-STARRS\,1 pixel mask was used to create a SExtractor weighting map excluding regions which fell on chip gaps or streaks from the removal of bright stars. We then extracted cutout images which were 5$\times$ the measured FWHM (available in the image headers) for each source on each image. We used the area at a distance of 4--5$\times$ FWHM to define our sky annulus, setting the sky value to be the median in this region. After subtracting the sky we measured the image parameters described in \cite{Terziev2013} and \cite{Hoekstra2005}, namely $e_{\nu}$, $p_{\nu}$ and $P_{\nu\nu}$ in a region within 4$\times$FWHM of the SExtractor reported centroid. Our weighting function $W$ was a Gaussian centred on the position of the object with a standard deviation of FWHM/2.35. This is simply the width of the seeing PSF and is substantially narrower than the weighting functions used for extragalactic weak lensing surveys as galaxies have significantly more flux away from their core than stars. This value was chosen as it was found that a larger choice meant that low significance detections were vulnerable to sky noise away from the object's position. The number of pixels in the corresponding region of the pixel mask image which were flagged as gaps or streaks were also counted. This process was repeated so we had positions, S/N estimates, image flags and shape parameters for each object on each image. 

On each image we then selected reference stars which had a S/N$>$10, had no SExtractor flags set, did not fall in gaps or streaks on the image, was classified by SExtractor as having an $>$80\% probability of being a star, and which had values of $p_1$ and $p_2$ which were less than 3 standard deviations from the median $p_1$ and $p_2$ images on that particular image. This last cut was to prevent objects with very high $p_{\nu}$ values from dominating our estimates of distortion across the image. Each image was required to have 10 or more reference stars meeting these conditions otherwise it was not used in subsequent analysis. For each reference star we measured the $p_1$ and $p_2$ parameters. We then fitted a second-order 2-D polynomial for both $p_1$ and $p_2$ using the MPFIT2DFUN routine \cite{Markwardt2009}. We found that for larger images third-order terms also become significant.

\begin{equation}
p_{\nu,smooth} = c_{\nu,0}  +c_{\nu,1}y +c_{\nu,2}x + c_{\nu,3}y^2 +c_{\nu,4}xy+c_{\nu,5}x^2
\end{equation}

We then used Equation~\ref{e_cor} to remove most of the image anisotropy. We do not do the final correction for the $\alpha$ parameter introduced by \cite{Hoekstra2005} at this stage as this is a {\it per object} rather than a {\it per image} parameter.

The detections for our target object across each image were then collected together and a fit for the variation with seeing and the $\alpha$ correction were applied as,

\begin{equation}
e_{fit,\nu}=c_{fit,0}+c_{fit,1}FWHM +{c_{fit,2}FWHM^2}+\alpha \sqrt{p_1^2+p_2^2}
\end{equation}

In doing this fit we exclude any detection with an ellipticity greater than 1, has $\sqrt{p_1^2+p_2^2}>3$, has a chip gap or streak within 4$\times$FWHM or which is detected at a S/N less than 15. This latter cut is more stringent than our cut for reference stars and is included to prevent individual, low-significance detections in good seeing from adversely affecting our fits. \cite{Hoekstra2005} suggest that the typical value of $\alpha$ is around 0.6 times the ellipticity at the reference seeing value for the object. To prevent the value of $\alpha$ becoming too large we carried out an initial fit where $\alpha$ was set to zero but $c_{fit,0}$, $c_{fit,1}$ and $c_{fit,2}$ where allowed any value. We  then determined $e_{fit,\nu}$ at a reference seeing of 1\arcsec\, from our initial fit ($e_{fit,\nu,ref}$) and then undertook a second fit where we limited $\alpha$ to be $-e_{fit,\nu,ref}<\alpha<e_{fit,\nu,ref}$ again also fitting for $c_{fit,0}$, $c_{fit,1}$ and $c_{fit,2}$. Once this was done we recalculated the value of $e_{fit,\nu,ref}$ for our new fit and used it as a diagnostic for stellar binarity. Only objects with six or more detections were considered for subsequent analysis. We found this threshold was the minimum possible before we became swamped with false detections from poor fits.

\subsection{Results of the Terziev test sample}
\begin{figure}
 \setlength{\unitlength}{1mm}
 \begin{tabular}{cc}
 \begin{picture}(70,100)
 \includegraphics{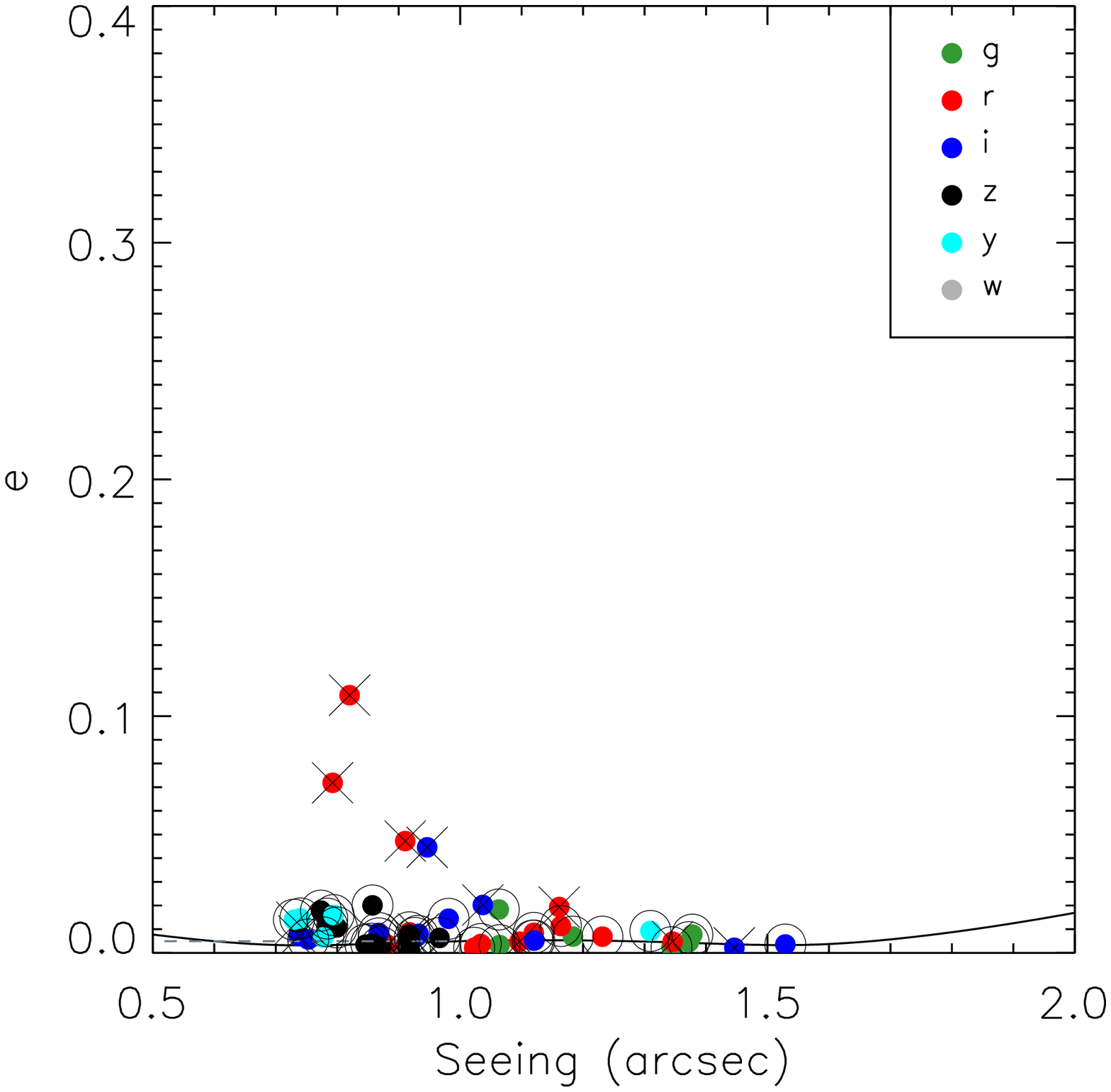}
 \end{picture}&
 \begin{picture}(70,100)
 \includegraphics{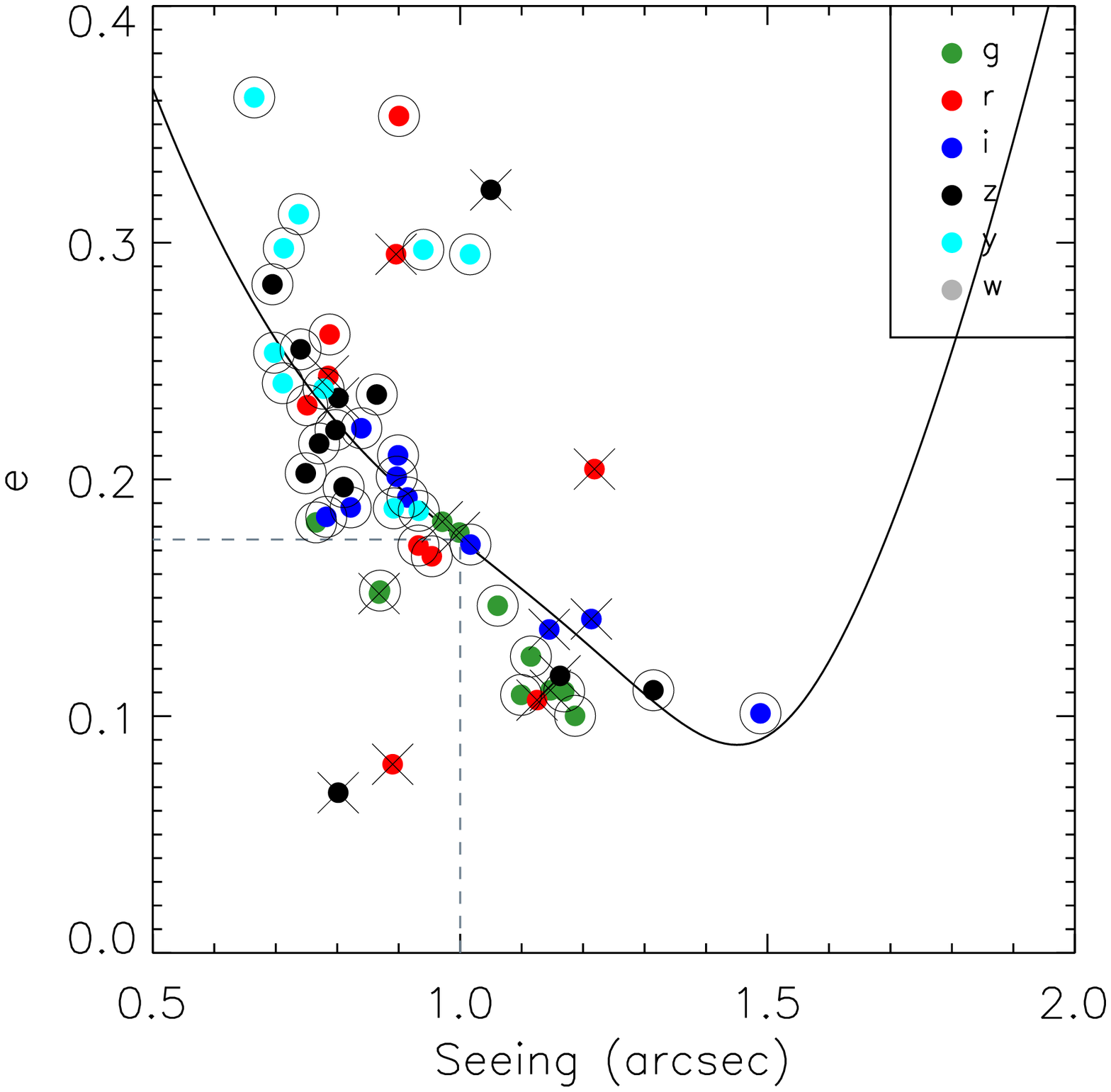}
 \end{picture}
  \end{tabular}
 \caption[]{Left: the single star PTF22.415, right: the 0.84 arcsec binary PTF23.553. Points with circles around them were images used to for the fits for each object. Points with crosses through them were excluded due to chip gaps, streaks due to bright stars or high polarisability ($p>3$). The dashed line marks the predicted ellipticity at a seeing of one arcsecond.} 
  \label{terziev_example}
 \end{figure}

\begin{figure}
 \setlength{\unitlength}{1mm}
 \begin{picture}(100,100)
 \includegraphics{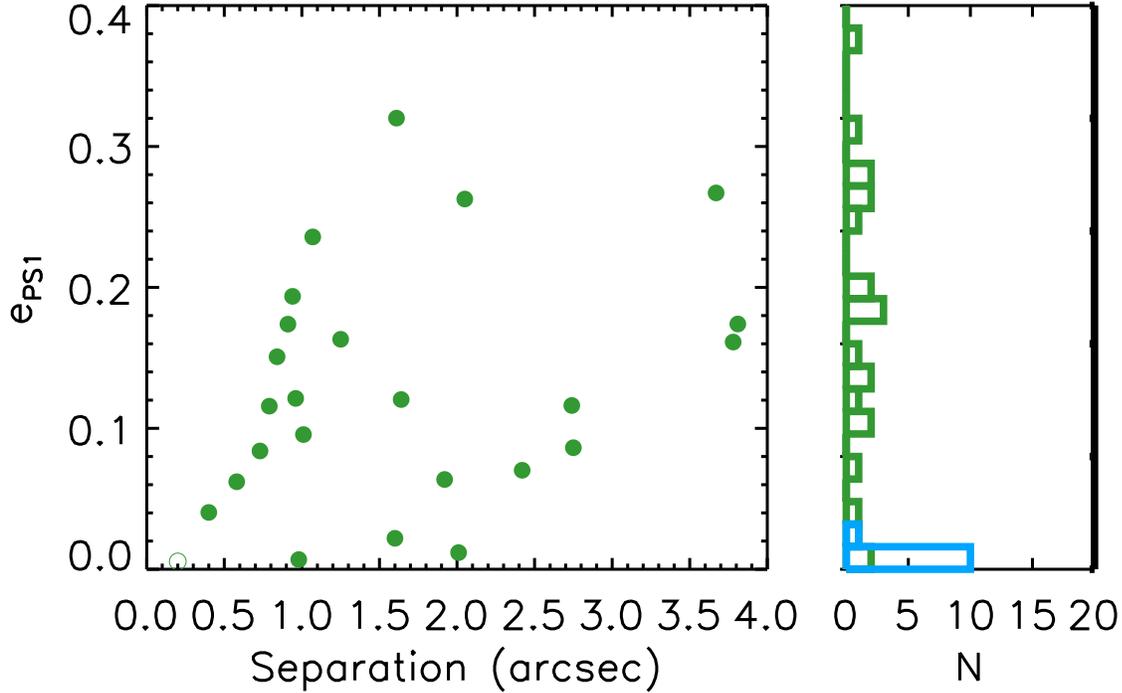}
 \end{picture}
 \caption[]{Left: the measured ellipticites at our reference seeing of 1\arcsec\, for binaries in the test sample of \protect\cite{Terziev2013}. The open circle is an object with a very small separation which Terziev notes would not have affected the ellipticity measurement. Right: a histogram of these ellipticities for binaries (green) and single stars (blue). Note the one binary at a separation of 2\arcsec\, which has a low elliipticity. This has the smallest flux ratio in the sample at 0.04.} 
  \label{terziev_test}
 \end{figure}

Forty of the 44 objects in the Terziev test sample had a sufficient number of measurements for fits to be performed. Of these, 12 were known single stars and 28 binaries (with one having a separation which \citealt{Terziev2013} notes is too small to affect the ellipticity measurements). Figure~\ref{terziev_example} shows two objects, one a single star and one a 0.84\arcsec\, binary. It is clear that the binary has both a higher ellipicity at the reference seeing and rising ellipticity as image quality improves. Figure~\ref{terziev_test} shows the ellipticities at our reference seeing of 1\arcsec\, for these 40 objects from the \cite{Terziev2013} sample. All but two binaries have ellipticities higher than 0.02 (the cutoff value Terziev suggests for PTF). Of these two binaries, one has a very small separation and the other has a flux ratio of 0.04, the smallest in the Terziev sample. Terziev's measurement of the ellipticity of this object was similarly low at 0.014. This shows that we are able to reliably recover known binaries in the the Terziev sample without producing a large number of false positives.

\cite{Terziev2013} use the absolute magnitude of the slope of the ellipiticity/seeing relation divided by the reference ellipticity as a measure of binary separation. We found a correlation between this statistic and binary separation but the scatter on the relation between the two was too large for it to act as a useful diagnostic.

\subsection{Testing with simulated data}
\begin{figure}
 \setlength{\unitlength}{1mm}
 \begin{picture}(100,100)
 \includegraphics{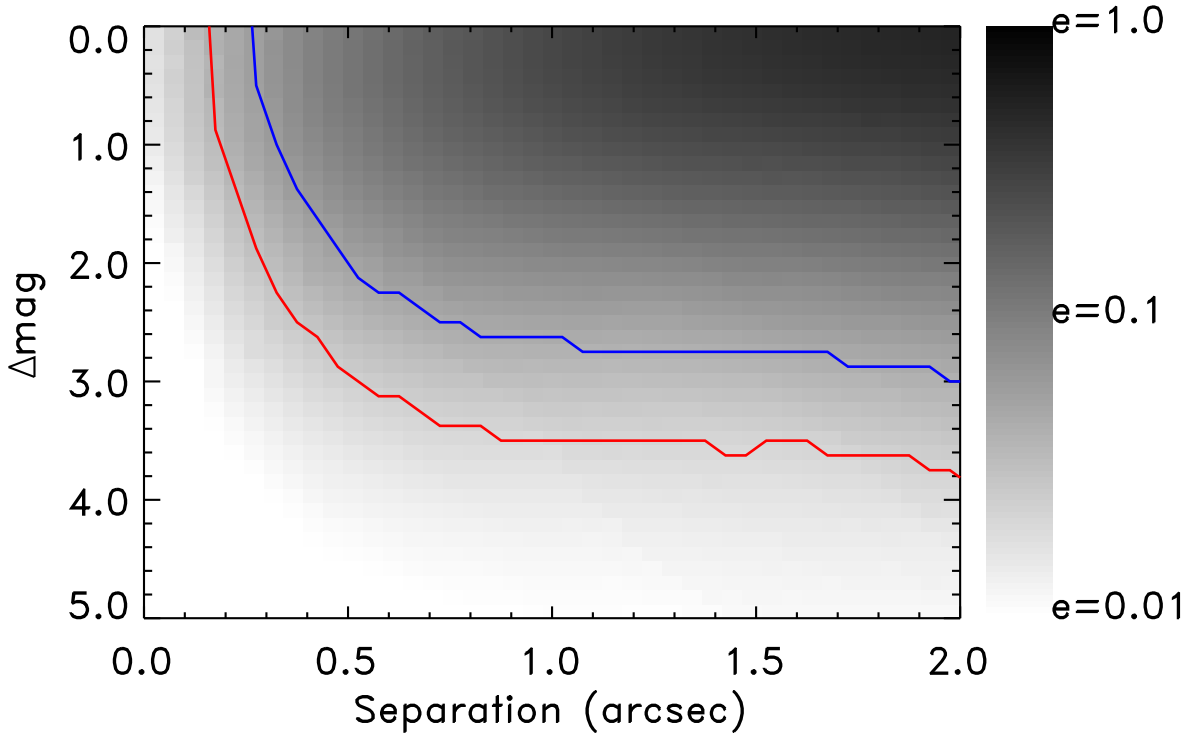}
 \end{picture}
 \caption[]{The measured ellipticities of a series of simulated binaries with varying flux ratio and separation. The red line marks the flux ratio at each separation where the measured ellipticity of the binary is above 0.02 (our chosen limit) and the blue line marks a more conservative $e=0.04$ limit.} 
  \label{terziev_sim}
 \end{figure}

To test our ability to probe binaries with different flux ratios and separations, we performed a series of simulations. We selected one single, low ellipiticity star from the Terziev sample and injected a copy of it at different separations and flux ratios. The seeing of the image chosen was 0.97\arcsec\,, close to our reference seeing of 1\arcsec\, We then ran our shape measurement code on these simulated binaries to determine the ellipticities across our parameter range. The resulting ellipticities are shown in Figure~\ref{terziev_sim} suggesting that binaries wider than 0.4\arcsec\, and with magnitude differences less than 3 will be readily identifiable with our method.

\begin{figure}
 \setlength{\unitlength}{1mm}
 \begin{picture}(100,120)
 \includegraphics{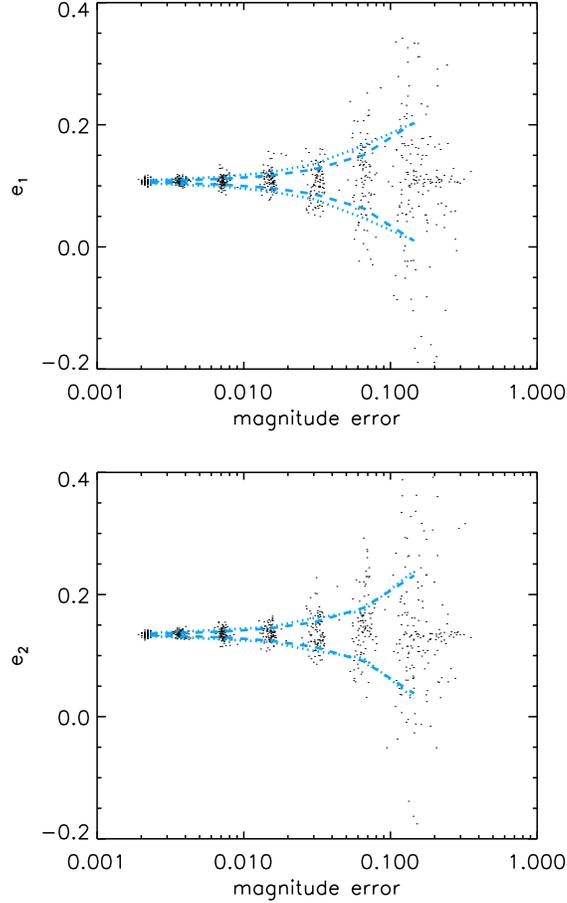}
 \end{picture}
 \caption[]{The individual ellipticity measurements, uncorrected for image anisotropy, for the same stellar image with different photon noise levels. The dotted line represents a robust (1.48 median square deviations) measure of the spread of ellipticities  in each magnitude error bin and the dashed line represents our $\sigma_{e} = \frac{2}{3}\sigma_{mag}$ fit.} 
  \label{terziev_noise}
 \end{figure}

To test the accuracy of our ellipticity measurements we took a well-exposed stellar image with seeing of approximately 1\arcsec, multiplied it by a factor less than one to reduce its brightness and added Poisson noise to it. This was then injected onto a blank region of a Pan-STARRS\,1 image. We repeated this process with different multiplying factors to produce a series of objects with a range of detection significances (and hence different errors on the measured magnitude of the star in SExtactor). We then ran each simulated image through our process and noted the measured magnitude error and ellipticities for each simulated object. Figure~\ref{terziev_noise} shows how the range of measured ellipticities increases with increasing magnitude error. We estimated the error on the ellipticity by dividing the data into seven bins and finding in each the median absolute deviation from the overall median value of each ellipticity parameter. We use this technique to prevent our error estimates being driven by a handful of outliers. This is median absolute deviation was multiplied by a factor of 1.48 to produce a robust estimate of the typical standard deviation of ellipticity measurements. We found that the error in each bin was well fitted by, 

\begin{equation}
\label{err_eq}
\sigma_e = \frac{2}{3}\sigma_{mag}
\end{equation}

and have used this form in the rest of our analysis.

\section{Searching for binary brown dwarfs}
We applied our binary search technique to a sample of ultracool dwarfs in the solar neighbourhood. To do this we searched a list of objects later than M6 compiled by W. Best (see Best et al., 2016, in prep for details) for objects with distance estimates less than 30\,pc and a total Pan-STARRS\,1 signal to noise ratio over a all epochs in either $z_{P1}$ or $y_{P1}$ greater than 15. This left us with 664 objects, all north of $\delta=-30^{\circ}$. We extracted postage stamp images from the Pan-STARRS\,1 postage stamp server for all objects. These images were from the PV3 processing run, were "warp" images (i.e. resampled on a regular grid of pixels aligned with the R.A. and Dec. axes) and covered 10\arcmin$\times$10\arcmin\, each. As we used individual epoch "warp" images calculating stellar centroid positions on each we are not affected by centroid offsets caused by proper motion. We also downloaded the corresponding mask images to allow us to flag objects which lay close to a chip gap or image streak. On each image we then ran SExtractor and followed our previously described reference star selection technique. We then corrected the ellipticity parameters for every object on the image using our smoothed estimate of the image anisotropy (using Equations~\ref{e_cor} and \ref{p_smooth}). 

For each of our targets we selected any individual epoch detection with S/N$>$15. We then fitted second-order polynomials to both the corrected values of $e_1$ and $e_2$ using a least-squares fitting method where the errors on each ellipticity measurements were determined using Equation~\ref{err_eq}. We then estimated the total ellipticity value at a seeing of 1\arcsec\, and propagated the errors from our covariance matrix to determine a standard deviation for this value. 

Out of a total of 664 objects in our input sample, 282 had six or more images with a sufficient number of reference stars, where the target was detected with sufficient significance and where the target was not affected by chip gaps or streaks. We then selected objects with a reference ellipticity greater than 0.02 and with that ellipticity being more significant than 4$\sigma$. Table~\ref{cands} lists these 27 objects and our evaluation of them. Table~\ref{duds} lists objects which did not pass our ellipticity cuts either due to a low ellipticity measurement or a low measurement significance.
\begin{table*}
\begin{minipage}{170mm}
\caption{A list of objects detected with an ellipticity at the reference seeing greater than 0.02 and more significant than 4$\sigma$. The column "Good" states if we believe this to be a believable fit or not, "Fig." shows the figure number the object appears in and "Known" lists if the object is a previously known binary.}

\label{cands}
\begin{center}
\footnotesize
\begin{tabular}{lccccl}
\hline
Name&$e_{PS1}$&Good&Fig.&Known&Notes\\
\hline
2MASSW J1421314+182740&0.023$\pm$0.002&N&&N&Affected by nearby star\\
2MASSI J1426316+155701&0.023$\pm$0.001&Y&\ref{bins1}&Y&Recovery of known binary\\
SIPS J1632$-$0631&0.024$\pm$0.002&N&&N&No consistent elliptcity\\
2MASSI J0835425$-$081923&0.025$\pm$0.002&N&&N&Strongly affected by one outlier measurement\\
2MASS J17343053$-$1151388&0.031$\pm$0.001&N&&N&Strongly affected by one outlier measurement\\
2MASS J17461199+5034036&0.035$\pm$0.008&N&&N&Strongly affected by one outlier measurement\\
2MASS J18000116$-$1559235&0.036$\pm$0.003&N&&N&Crowded field\\
Kelu-1&0.041$\pm$0.003&Y&\ref{bins1}&Y&Recovery of known binary\\
LHS 2930&0.042$\pm$0.001&N&&N&Saturated\\
2MASSW J2206228$-$204705&0.042$\pm$0.001&Y&\ref{bins2}&Y&Recovery of known binary\\
2MASS J17312974+2721233&0.047$\pm$0.001&N&&N&Affected by nearby star\\
2MASS J09153413+0422045&0.052$\pm$0.006&Y&\ref{bins2}&Y&Recovery of known binary\\
2MASS J05301261+6253254&0.057$\pm$0.006&N&&N&Strongly affected by one outlier measurement\\
WISEPA J061135.13$-$041024.0&0.063$\pm$0.016&Y&\ref{bins3}&Y&Recovery of known binary\\
WISE J180952.53$-$044812.5&0.065$\pm$0.009&Y&\ref{bins6}&N&Discovery\\
2MASS J17072343$-$0558249&0.065$\pm$0.001&Y&\ref{bins3}&Y&Recovery of known binary\\
2MASS J05431887+6422528&0.072$\pm$0.003&Y&\ref{bins6}&N&Discovery\\
2MASS J11000965+4957470&0.099$\pm$0.015&N&&N&Extremely noisy detections CHECK\\
SIMP J1619275+031350&0.125$\pm$0.025&Y&\ref{bins4}&Y&Recovery of known binary\\
LP 44-334&0.130$\pm$0.001&Y&\ref{bins6a}&N&Discovery\\
WISE J072003.20$-$084651.2&0.138$\pm$0.001&N&&N&Saturated\\
2MASS J19303829$-$1335083&0.141$\pm$0.001&N&&N&Saturated\\
DENIS-P J220002.05$-$303832.9&0.214$\pm$0.008&Y&\ref{bins4}&Y&Recovery of known binary\\
2MASS J15500845+1455180&0.231$\pm$0.011&Y&\ref{bins5}&Y&Recovery of known binary\\
G 196-3B&0.317$\pm$0.027&N&&N&Affected by nearby star\\
LP 412-31&0.385$\pm$0.001&N&&N&Strongly affected by one outlier measurement\\
DENIS J020529.0$-$115925&0.785$\pm$0.160&N&\ref{bins5}&Y&Strongly affected by one outlier measurement\\
\hline
\normalsize
\end{tabular}
\end{center}
\end{minipage}
\end{table*}

\begin{table*}
\begin{minipage}{170mm}
\caption{A list of objects which did not have ellipticity at the reference seeing greater than 0.02 and more significant than 4$\sigma$. A full version of this table will be available online}

\label{duds}
\begin{center}
\footnotesize
\begin{tabular}{lccccc}
\hline
Name&R.A.&Dec.&$e_{PS1}$&$e_{1}$&$e_{2}$\\
\hline
SDSS J000112.18+153535.5&00:01:12.28&+15:35:33.7&0.338$\pm$0.111&$-$0.181$\pm$0.111&0.286$\pm$0.111\\
2MASSW J0015447+351603&00:15:44.82&+35:15:59.9&0.005$\pm$0.002&0.003$\pm$0.003&0.005$\pm$0.002\\
2MASS J00192626+4614078&00:19:26.40&+46:14:06.8&0.001$\pm$0.0&0.0$\pm$0.0&0.0$\pm$0.0\\
BRI 0021$-$0214&00:24:24.58&$-$01:58:18.4&0.006$\pm$0.001&$-$0.001$\pm$0.0&$-$0.005$\pm$0.001\\
LP 349$-$25&00:27:56.31&+22:19:30.6&0.002$\pm$0.0&0.001$\pm$0.0&$-$0.002$\pm$0.0\\
PSO J007.9194+33.5961&00:31:40.65&+33:35:45.9&0.133$\pm$0.099&$-$0.08$\pm$0.099&$-$0.106$\pm$0.099\\
\hline
\normalsize
\end{tabular}
\end{center}
\end{minipage}
\end{table*}

\begin{figure}
 \setlength{\unitlength}{1mm}
 \begin{tabular}{cc}
 \begin{picture}(40,170)
 \includegraphics{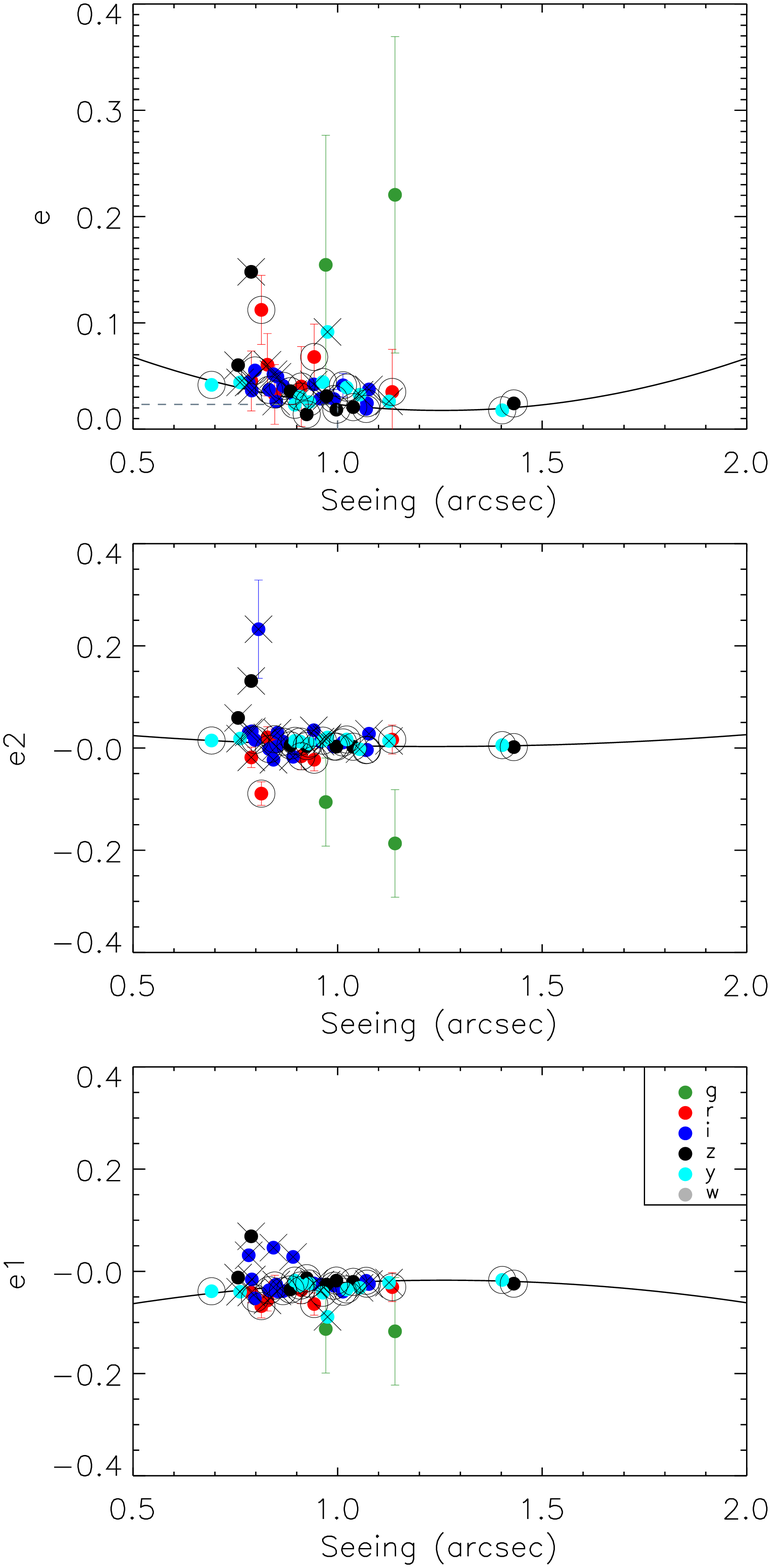}
 \end{picture}&
 
 \begin{picture}(40,170)
 \includegraphics{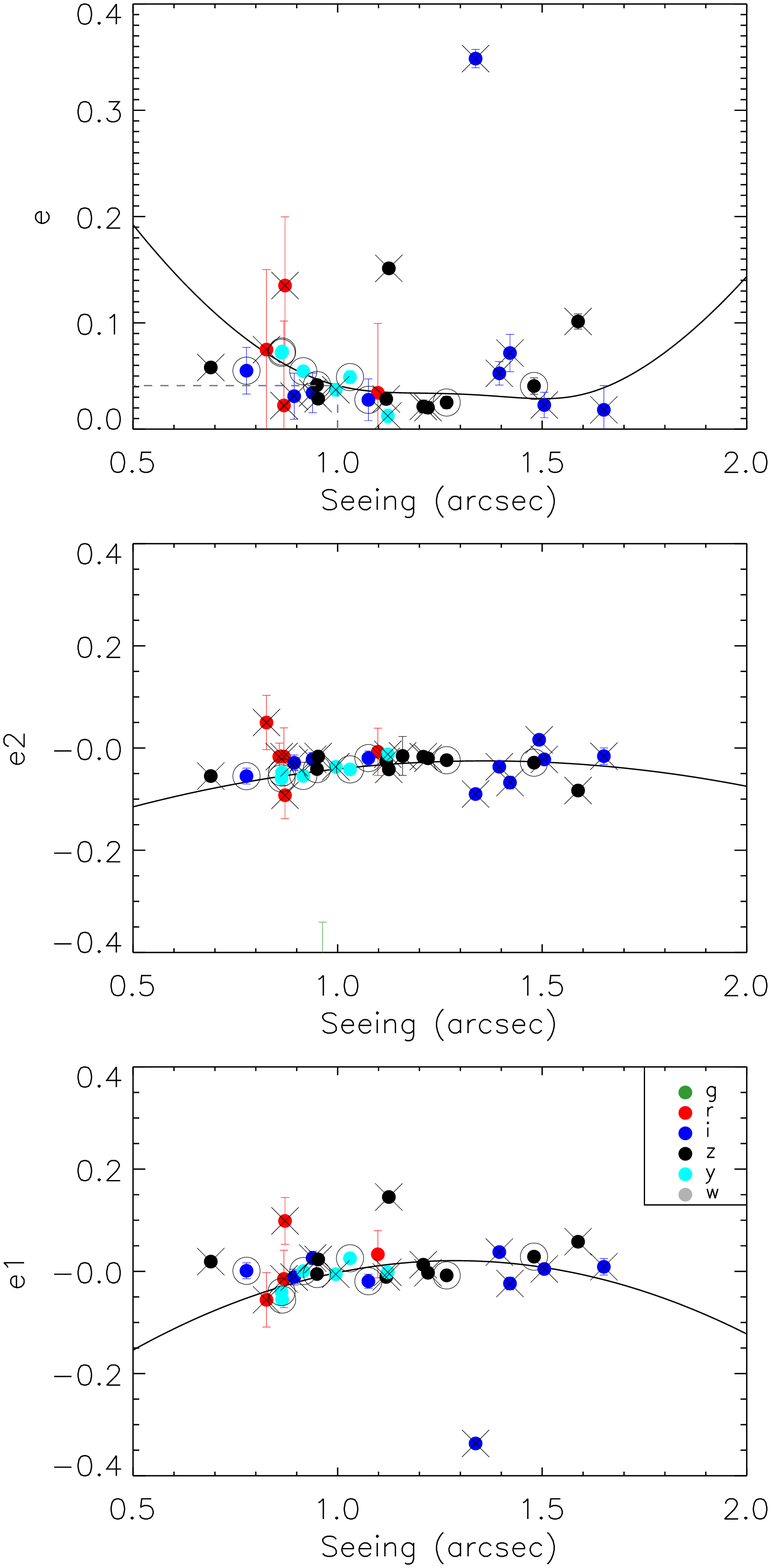}
 \end{picture}
 \end{tabular}
 \caption[]{Ellipticity measurements and fits for the known binaries 2MASSI J1426316+155701 (left) and Kelu-1 (right). The top plot shows the total ellipticity, the middle plot the "$\times$" polarisation $e_2$ and the bottom plot the "+" polarisation $e_2$. Points used in the 2nd order polynomial fits are outlined by a circle and those rejected for data quality reasons are crossed out. Note due to the coordinate system of the postage stamps used a positive $e_2$ is an elongation in the NW-SE direction and a positive $e_1$ is an elongation along the $x$ (i.e. R.A.) axis.} 
  \label{bins1}
 \end{figure}
 
\begin{figure}
 \setlength{\unitlength}{1mm}
 \begin{tabular}{cc}
 \begin{picture}(40,170)
 \includegraphics{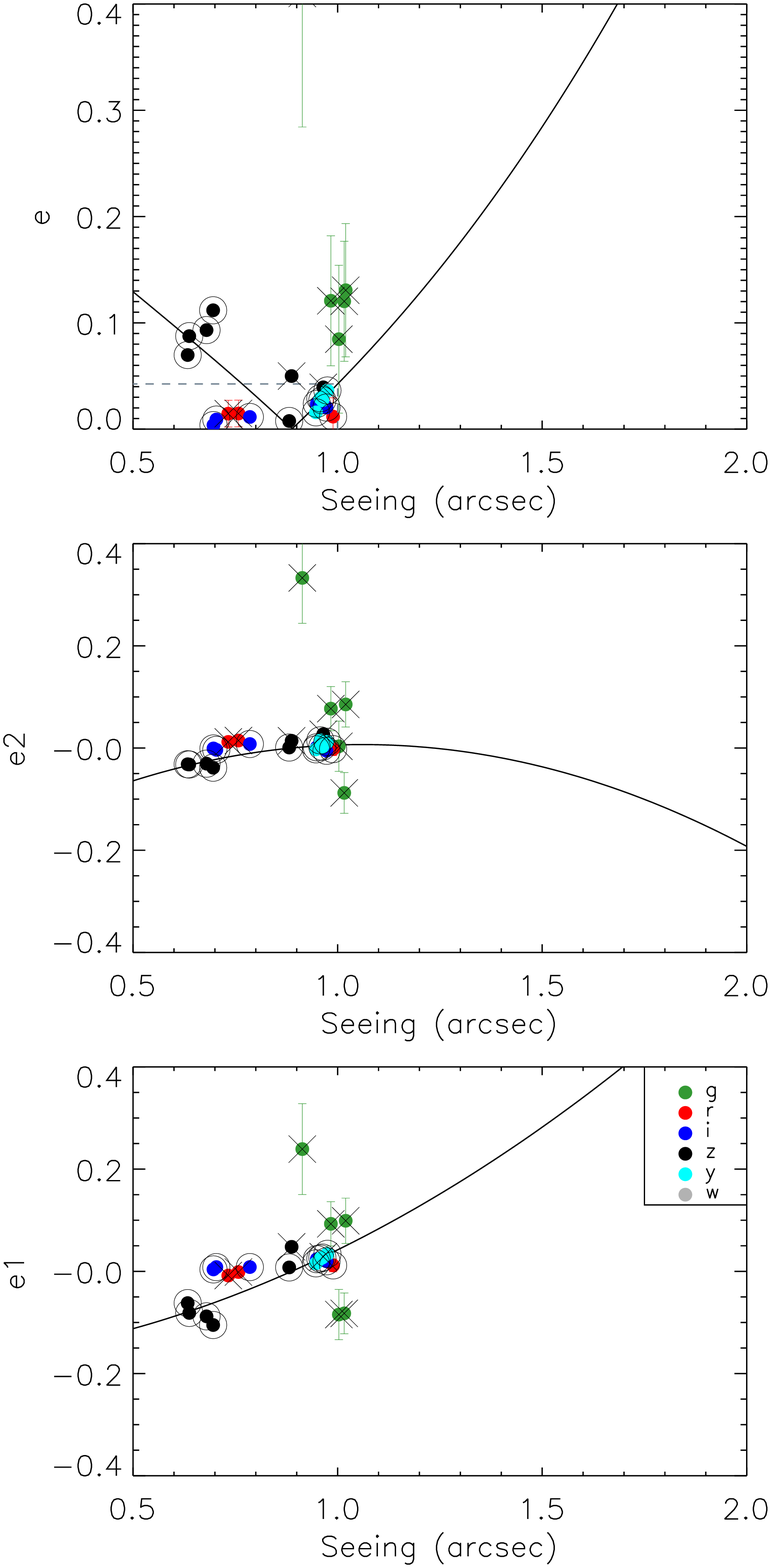}
 \end{picture}&
 \begin{picture}(40,170)
 \includegraphics{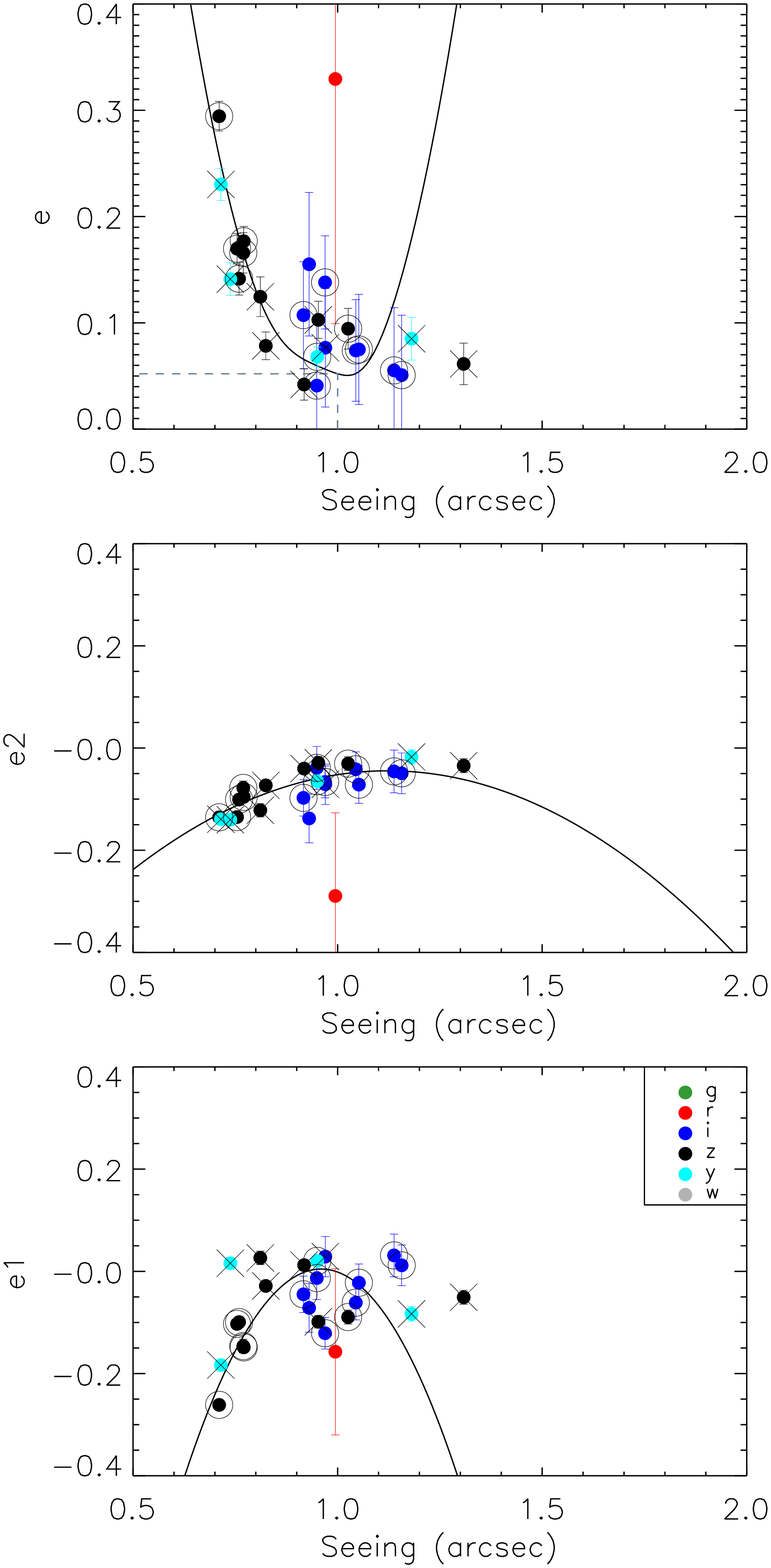}
 \end{picture}
 \end{tabular}
 \caption[]{Ellipticity measurements and fits for the known binaries 2MASSW J2206228$-$204705 (left) and 2MASS J09153413+0422045 (right). The top plot shows the total ellipticity, the middle plot the "x" polarisation $e_2$ and the bottom plot the "+" polarisation $e_2$. Points used in the 2nd order polynomial fits are outlined by a circle and those rejected for data quality reasons are crossed out. Note due to the coordinate system of the postage stamps used a positive $e_2$ is an elongation in the NW-SE direction and a positive $e_1$ is an elongation along the $x$ (i.e. R.A.) axis.} 
  \label{bins2}
 \end{figure}
 
 \begin{figure}
 \setlength{\unitlength}{1mm}
 \begin{tabular}{cc}
 \begin{picture}(40,170)
 \includegraphics{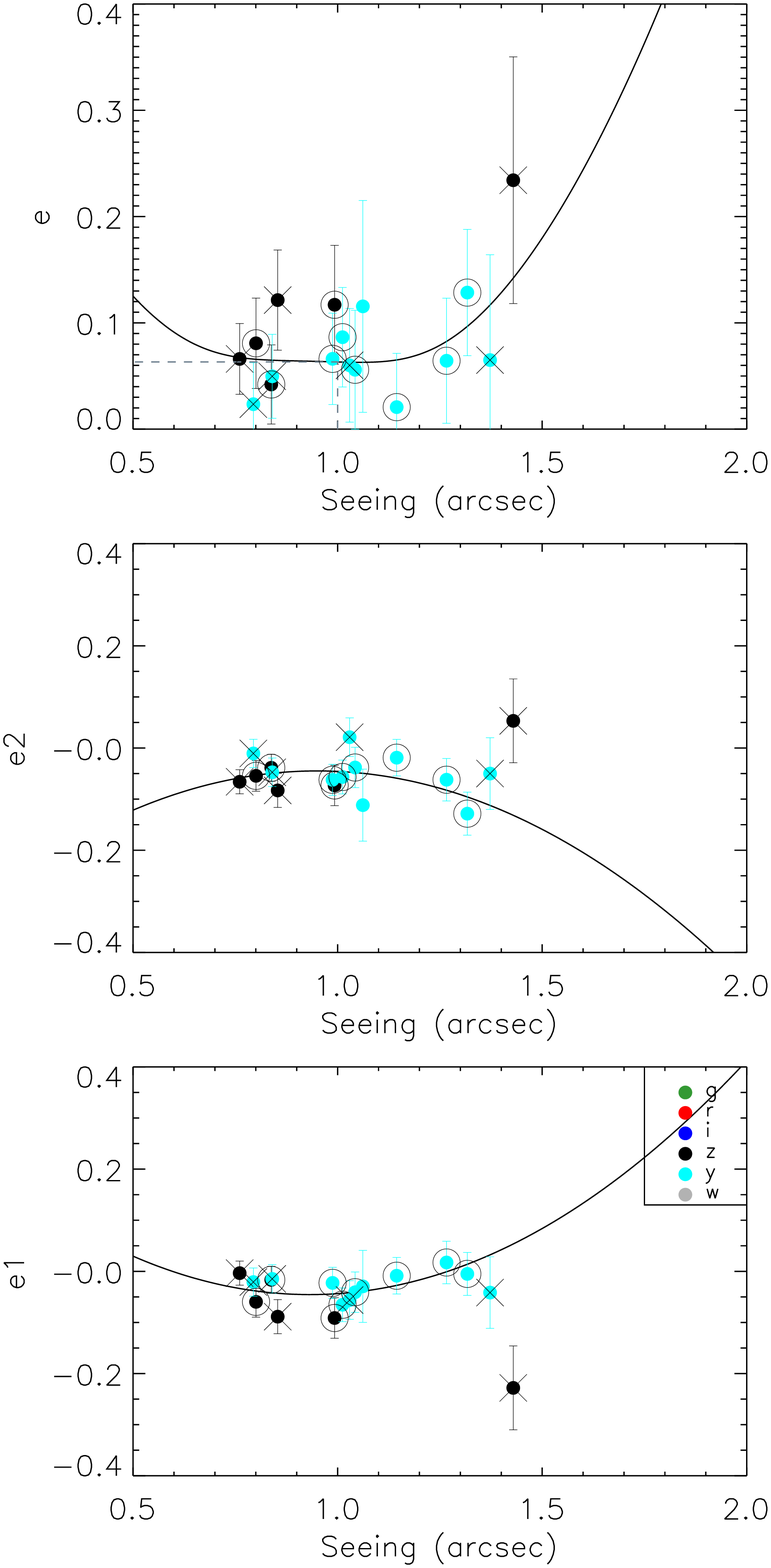}
 \end{picture}&
  \begin{picture}(40,170)
 \includegraphics{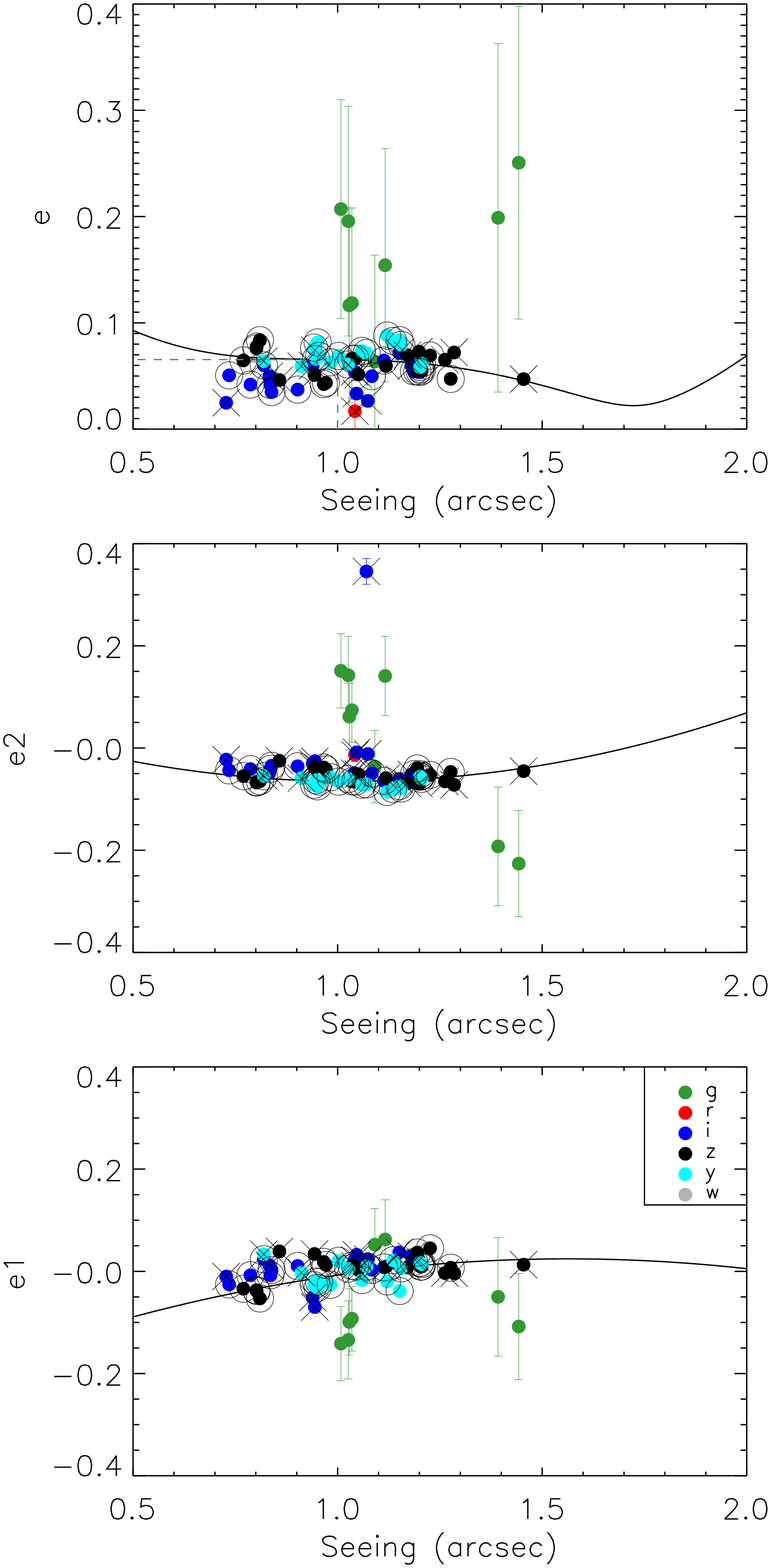}
 \end{picture}
 \end{tabular}
 \caption[]{Ellipticity measurements and fits for the known binaries WISEPA J061135.13$-$041024.0 (left) and 2MASS J17072343$-$055824 (right). The top plot shows the total ellipticity, the middle plot the "x" polarisation $e_2$ and the bottom plot the "+" polarisation $e_2$. Points used in the 2nd order polynomial fits are outlined by a circle and those rejected for data quality reasons are crossed out. Note due to the coordinate system of the postage stamps used a positive $e_2$ is an elongation in the NW-SE direction and a positive $e_1$ is an elongation along the $x$ (i.e. R.A.) axis.} 
  \label{bins3}
 \end{figure}
 
 \begin{figure}
 \setlength{\unitlength}{1mm}
 \begin{tabular}{cc}
 \begin{picture}(40,170)
 \includegraphics{./2M1426_e.ps}
 \end{picture}&
  \begin{picture}(40,170)
 \includegraphics{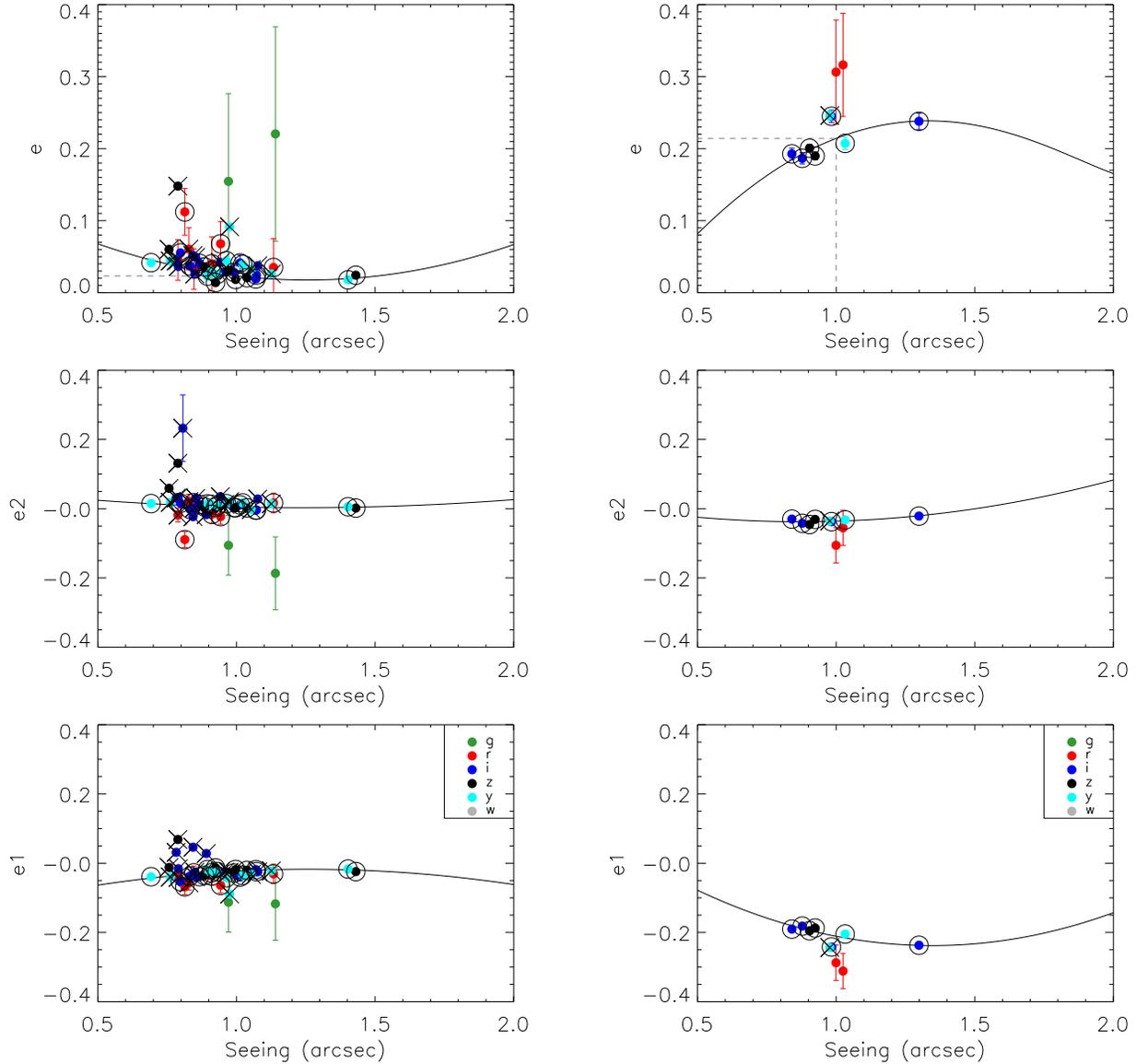}
 \end{picture}
 \end{tabular}
 \caption[]{Ellipticity measurements and fits for the known binaries SIMP J1619275+031350 (left) and DENIS-P J220002.05$-$303832.9 (right). The top plot shows the total ellipticity, the middle plot the "x" polarisation $e_2$ and the bottom plot the "+" polarisation $e_2$. Points used in the 2nd order polynomial fits are outlined by a circle and those rejected for data quality reasons are crossed out. Note due to the coordinate system of the postage stamps used a positive $e_2$ is an elongation in the NW-SE direction and a positive $e_1$ is an elongation along the $x$ (i.e. R.A.) axis.} 
  \label{bins4}
 \end{figure}

 \begin{figure}
 \setlength{\unitlength}{1mm}
 \begin{tabular}{cc}
 \begin{picture}(40,170)
 \includegraphics{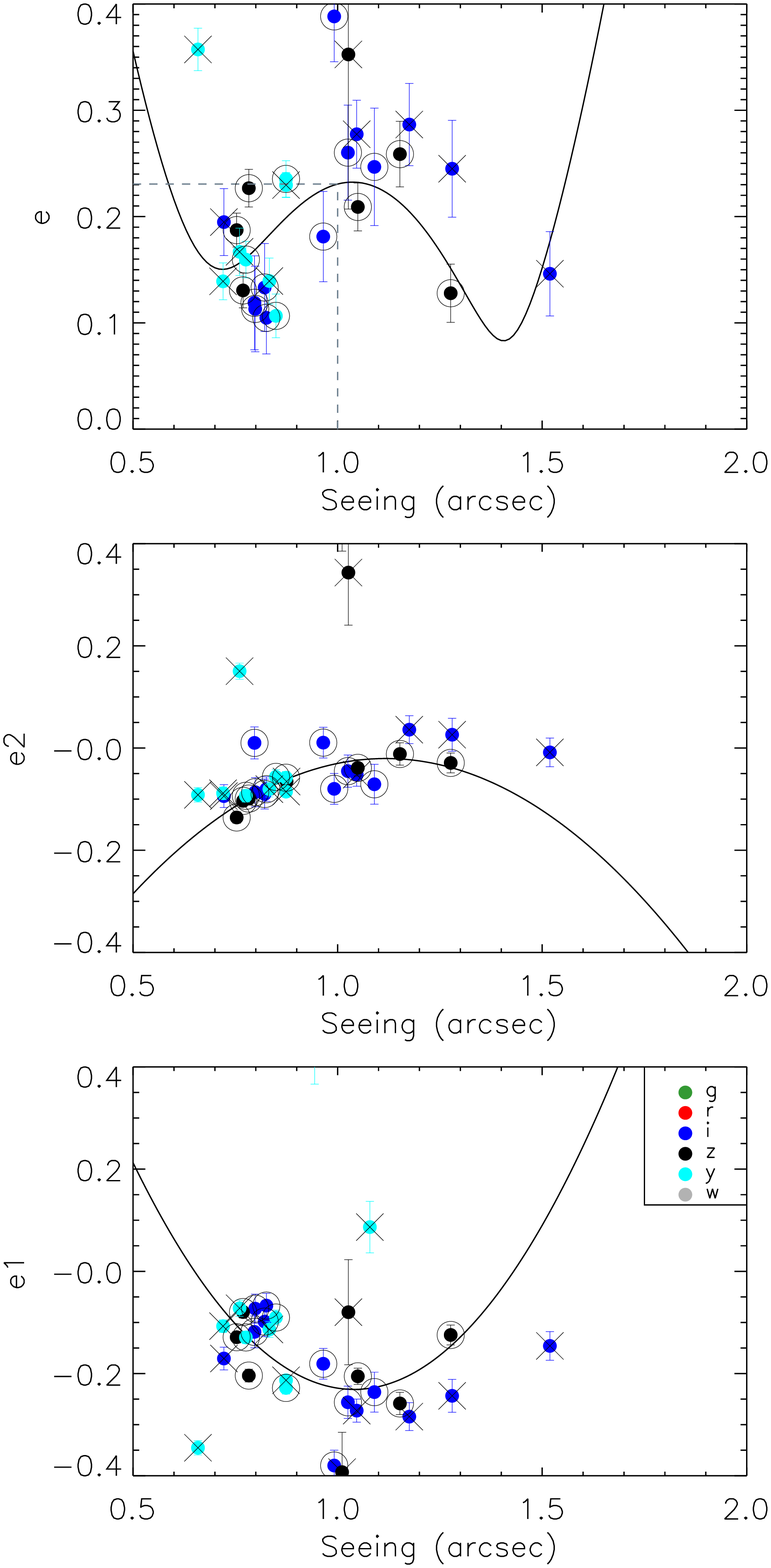}
 \end{picture}&
  \begin{picture}(40,170)
 \includegraphics{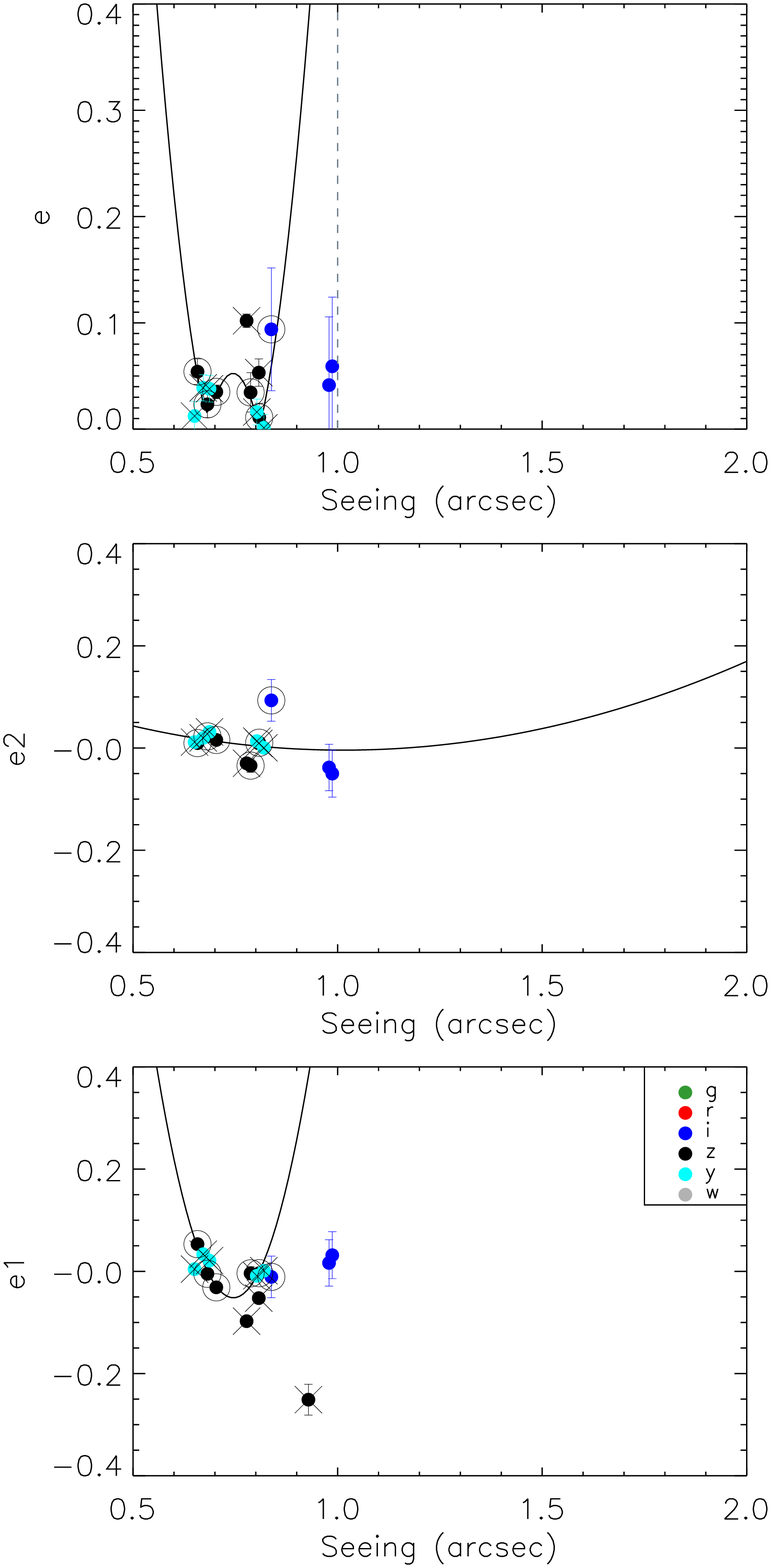}
 \end{picture}
 \end{tabular}
 \caption[]{Ellipticity measurements and fits for the known binaries 2MASS J15500845+1455180 (left) and DENIS J020529.0$-$115925 (right). The top plot shows the total ellipticity, the middle plot the "x" polarisation $e_2$ and the bottom plot the "+" polarisation $e_2$. Points used in the 2nd order polynomial fits are outlined by a circle and those rejected for data quality reasons are crossed out. Note due to the coordinate system of the postage stamps used a positive $e_2$ is an elongation in the NW-SE direction and a positive $e_1$ is an elongation along the $x$ (i.e. R.A.) axis.} 
  \label{bins5}
 \end{figure}

 \begin{figure}
 \setlength{\unitlength}{1mm}
 \begin{tabular}{cc}
 \begin{picture}(40,170)
 \includegraphics{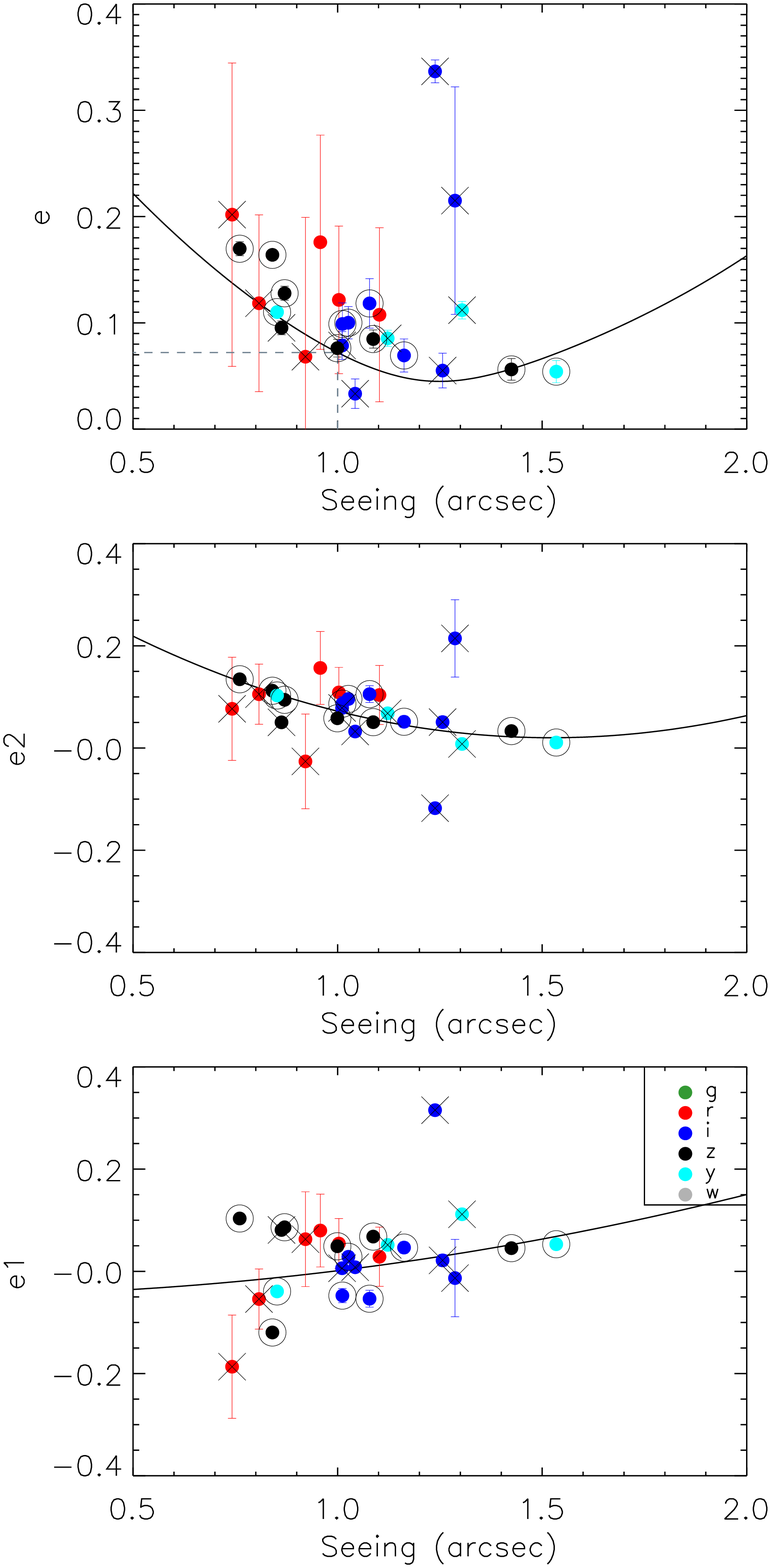}
 \end{picture}&
  \begin{picture}(40,170)
 \includegraphics{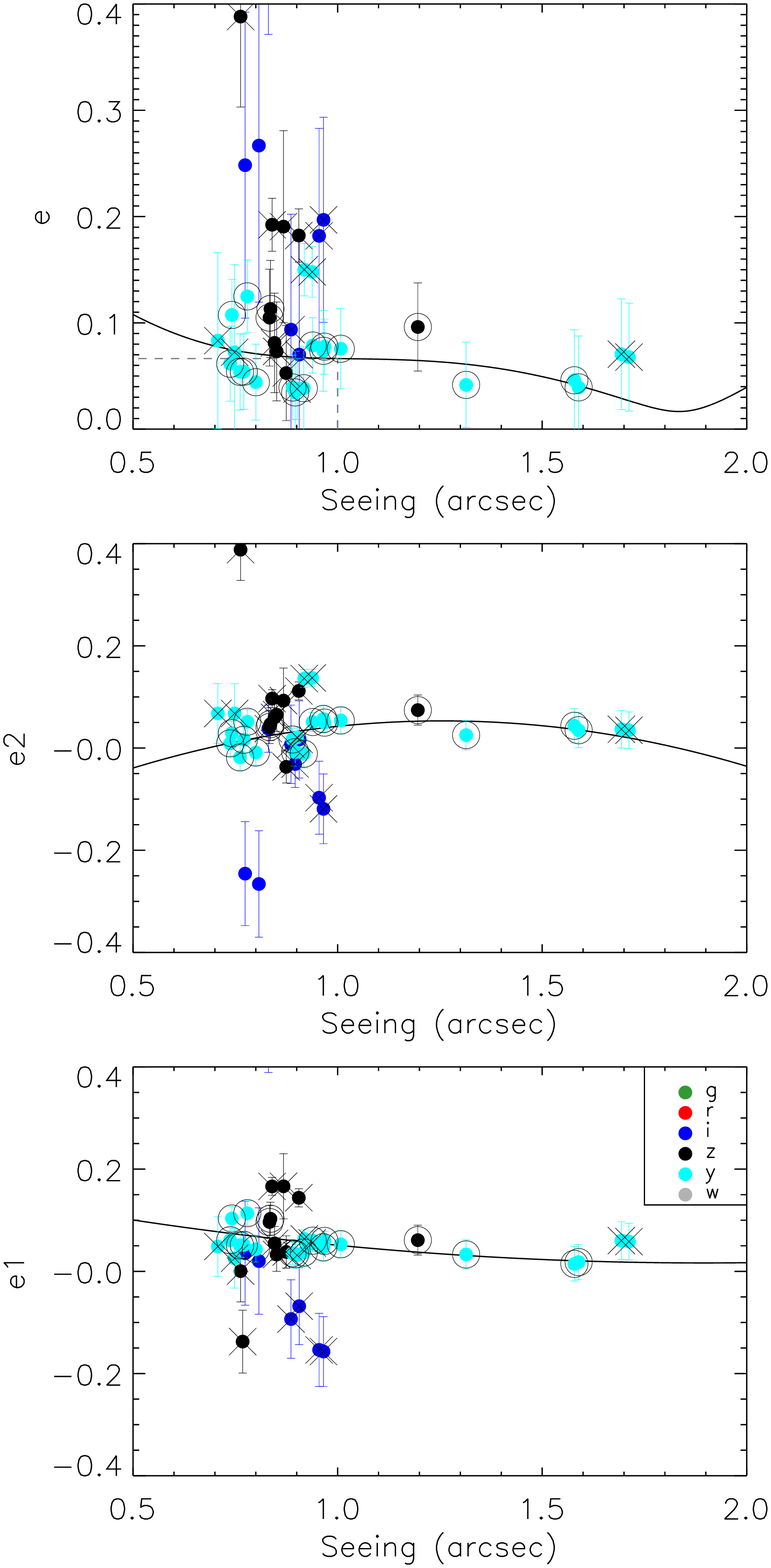}
 \end{picture}
 \end{tabular}
 \caption[]{Ellipticity measurements and fits for 2MASS J05431887+6422528 (left) and WISE J180952.53$-$044812.5 (right). The top plot shows the total ellipticity, the middle plot the "x" polarisation $e_2$ and the bottom plot the "+" polarisation $e_2$. Points used in the 2nd order polynomial fits are outlined by a circle and those rejected for data quality reasons are crossed out. Note due to the coordinate system of the postage stamps used a positive $e_2$ is an elongation in the NW-SE direction and a positive $e_1$ is an elongation along the $x$ (i.e. R.A.) axis.} 
  \label{bins6}
 \end{figure}

 \begin{figure}
 \setlength{\unitlength}{1mm}
  \begin{picture}(40,170)
 \includegraphics{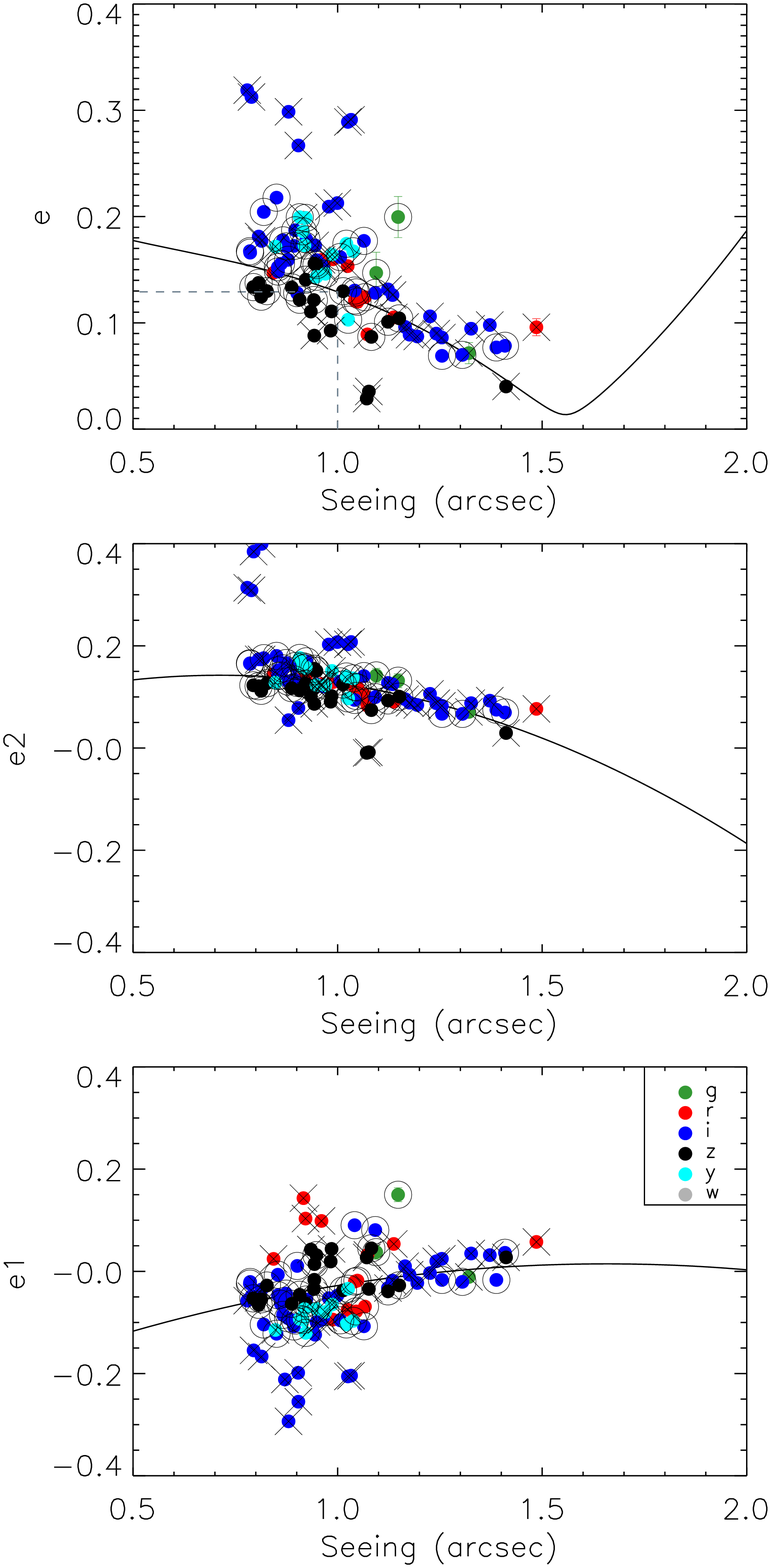}
 \end{picture}
 \caption[]{Ellipticity measurements and fits for LP 44-334. The top plot shows the total ellipticity, the middle plot the "x" polarisation $e_2$ and the bottom plot the "+" polarisation $e_2$. Points used in the 2nd order polynomial fits are outlined by a circle and those rejected for data quality reasons are crossed out. Note due to the coordinate system of the postage stamps used a positive $e_2$ is an elongation in the NW-SE direction and a positive $e_1$ is an elongation along the $x$ (i.e. R.A.) axis.} 
  \label{bins6a}
 \end{figure}

\subsection{Notes on previously known binaries}
In this section we discuss previously known binaries which we have recovered. Note that due to the pixel coordinates of our images (increasing $x$ coordinate with decreasing R.A., increasing $y$ with increasing Dec.), a positive value of $e_1$ will represent an elongation in the R.A. direction and a positive value of $e_2$ will be derived from an elongation in the north-east to south-west direction. This means that the right-hand side of the $e_2$ equation shown in Equation~\label{e_rot} is multiplied by a factor of $-1$. We summarise our alignment estimates and the literature position angles in Table~\ref{known_pa}.

\subsubsection{2MASSI J1426316+155701}
This M8.5+L1 binary was discovered by \cite{Close2003} having a separation of 0.152$\pm$0.006\arcsec\, and position angle of 344.1$\pm$0.7$^{\circ}$ in June 2001. Later observations by \cite{Konopacky2010} from March 2009 give a separation of 0.3226$\pm$0.0006$^{\arcsec}$ and position angle of 343.8$\pm$0.08$^{\circ}$, suggesting an edge-on orbit. We find our fit (see Figure~\ref{bins1}) is dominated by a negative $e_1$, suggesting an elongation along the North-South axis, in good agreement with the previous position angles.
\subsubsection{Kelu-1}
This L1.5-L3+L3-L4.5 binary was discovered by \cite{Liu2005} and has been monitored by M. Liu and T. Dupuy ever since (Dupuy \& Liu in prep.). During Pan-STARRS\,1 observations this binary would have had a separation of around 0.39\arcsec\, and a position angle of 255$^{\circ}$ (Dupuy et al., in prep.). Our fit shows a consistently negative $e_2$ which suggests an elongation in the NE-SW direction (see Figure~\ref{bins1}).
\subsubsection{2MASSW J2206228$-$204705}
Discovered by \cite{Close2003}, this M8+M8 binary has been monitored by M. Liu and T. Dupuy (Dupuy \& Liu in prep.) since its initial identification. Our fit (see Figure~\ref{bins2}) is heavily driven by four $z_{P1}$ observations taken on the 10th of November 2010 with very good seeing (0.6-0.7\arcsec\,). The main characteristic of the fit is a highly negative value of $e_1$ (elongation along the N-S axis) and a slightly negative value of $e_2$ (suggesting a tilt towards as NE-SW alignment. The orbital solution for 2MASSW J2206228$-$204705 suggests a separation of 0.139\arcsec\, and a position angle of 216$^{\circ}$ (Dupuy et al. in prep.), in agreement our suggested alignment. It should be noted that this binary is much tighter than we would normally expect to detect, but it appears that we are able to detect this is due to a single night of exceptional data.
\subsubsection{2MASS J09153413+0422045}
This L7+L7 binary was discovered by \cite{Reid2006a} as having a separation of 0.73\arcsec\, and position angle of 205$^{\circ}$. Our solution (see Figure~\ref{bins2}) results in a highly negative value of $e_1$ and a smaller but negative value of $e_2$. This is suggests a binary primarily elongated along the N-S axis but tilted towards the NE-SW direction, in agreement with \cite{Reid2006a}'s position angle.
\subsubsection{WISEPA J061135.13$-$041024.0}
Discovered by \cite{Gelino2014}, this L/T transition (L9+T1.5) binary has a separation of 0.384\arcsec\, and a position angle of 32.5$^{\circ}$. Our fit (see Figure~\ref{bins3}) is dominated by a negative value of $e_2$ (suggesting a NE-SW alignment) with a slightly negative value of $e_1$, suggesting a slight elongation in $y$. This is consistent with the known position angle of the binary.

\subsubsection{2MASS J17072343$-$055824}
This known M9+L3 \citep{Reid2006a} binary had a separation of 1.04 $\pm$0.04\arcsec,\, P.A.=145$\pm$2$^{\circ}$ in March 2003 \citep{McElwain2006}. This is significantly discrepant from our negative value of $e_2$, suggesting a position angle around 45$^{\circ}$ (see Figure~\ref{bins3}). An inspection of the Pan-STARRS\,1 images along with an image from the Vista Hemisphere Survey \citep{Cross2012} confirm that the binary is aligned with a position angle of roughly 45$^{\circ}$. \cite{McElwain2006}'s discovery image shows no additional object which could affect this measurement. Our position angle estimate is in much better agreement with the estimate of 35$^{\circ}$ by \cite{Reid2006a}.

\subsubsection{SIMP J1619275+031350}
\cite{Artigau2011} discovered this object was a 0.691$\pm$0.002\arcsec\, T2.5$+$T4.0 binary. The position angle is 71.23$\pm$0.23$^{\circ}$, in agreement with our positive value of $e_1$ and negative value of $e_2$ (see Figure~\ref{bins4}).

\subsubsection{DENIS-P J220002.05$-$303832.9}
This object was identified as a 1.094$\pm$0.06\arcsec\,, P.A.=176.7$\pm$2.0$^{\circ}$ M9+L0 by \cite{Burgasser2006a}. This position angle agrees well with our highly negative value of $e_1$ (see Figure~\ref{bins4}).
\subsubsection{2MASS J15500845+1455180}
This object was identified as a 0.91\arcsec\, L3.5$+$L4 binary by \cite{Burgasser2009a}. Our recovery of this system shows a negative value of $e_1$, suggesting an elongation along the N-S axis, consistent with \cite{Burgasser2009a}'s position angle of 16.6$\pm$1.3$^{\circ}$. We detect a slight negative $e_2$ at increasing seeing (see Figure~\ref{bins5}). This is because our $x$ pixel number increases with decreasing R.A., meaning the a negative $e_2$ suggests a binary tilted towards the north-east.

\subsubsection{DENIS J020529.0$-$115925}
This is a L7+L7 \citep{Reid2006a} binary detected by \cite{Bouy2003} with a separation of 0.287$\pm$0.005\arcsec\, and a position angle of 246$\pm$1$^{\circ}$ (see Figure~\ref{bins5}), however the reference ellipticity is anomolously high. 
This is the result of a conspiracy of circumstances: we have very few datapoints, all of them at significantly better seeing than 1\arcsec.\, This leads to the ellipticiity at the reference seeing being an extrapolation which is strongly affected by one poor datapoint with the worst seeing. We note that the rising $e_1$ at better seeing points to a binary elongated along the E-W axis.

\subsection{Newly identified binaries}

\subsubsection{WISE J180952.53$-$044812.5}
WISE J180952.53$-$044812.5 was discovered by \cite{Mace2013a} who classified it as T0.5. We identified this object as having a positive value of $e_1$ indicating an elongation in the R.A. axis (see Figure~\ref{bins6}). This was independently discovered as a binary by Best et al. (in prep.) with a separation of 0.3\arcsec\, and position angle of 112$^{\circ}$. This upcoming discovery paper measures relative photometry of $\Delta J$=$-$0.442$\pm$0.059\,mag and $\Delta K$=0.410$\pm$0.023\,mag from Keck-AO observations. Note that this is a flux reversal binary with the western component brighter in $J$ and the eastern component brighter in $K$. Based on these colours and the typical colours for ultracool dwarfs in Table~15 of \cite{Dupuy2012} we estimate that the eastern component has spectral type L8-L9 and the western component is a T2-T3. 
\subsubsection{2MASS J05431887+6422528}
This object was identified as an L1 by \cite{Reid2008}. We measured a significantly positive value of $e_2$ indicating an elongation in the NW-SE direction. We observed this object on 2014~Jan~22~UT using the NIRC2 imaging
camera on the Keck~II telescope on Mauna Kea, Hawaii.  We obtained 4
images in $K$ band and 5 images in $J$ band, using the wide camera
mode of NIRC2.  We reduced the images in a standard fashion using
custom IDL scripts. We constructed flat fields from the differences of
images of the telescope dome interior with and without lamp
illumination. We subtracted an average bias from the images and
divided by the flat-field. Then we created a master sky frame from the
median average of the bias-subtracted, flat-fielded images and
subtracted it from the individual reduced images. We registered and
stacked the individual reduced images to form a final mosaic.
We used the StarFinder package \citep{Diolaiti2000a} that iteratively solves for both the binary parameters and an empirical image of the point-spread function.
We determined the uncertainties in these binary parameters
from the rms scatter among each data set. 
To correct for non-linear
distortions in NIRC2, we used the calibration of (\citealt{Fu2012};
priv.\
comm.),\footnote{\url{http://astro.physics.uiowa.edu/~fu/idl/nirc2wide/}}
with a corresponding pixel scale of 39.686\,mas\,pixel$^{-1}$ and the
same $+0\fdg252\pm0\fdg0.009$ correction for the orientation given in
the NIRC2 image headers as in \cite{Yelda2010}. 
The photometric errors are computed as the rms of individual frames, which sometimes does not fully capture systematic errors, e.g., from PSF fitting.  For practical purposes, the flux ratios are more precise than the integrated-light photometry, which will be the limiting factor in the precision of the resolved photometry for the binary. The separation and
PA measured in $K$ band ($655.37\pm3.39$\,mas and
$320.14\pm0.24$\,deg) were consistent with and more precise than measured in $J$ band ($656.66\pm3.55$\,mas and $319.85\pm0.36$\,deg).
The flux ratio in $J$ band ($0.284\pm0.030$\,mag) is further from
unity than in $K$ band ($0.259\pm0.064$\,mag), implying that the
brighter, southeastern component is marginally bluer as expected. The integrated spectral type of L1 \citep{Reid2008} and the small magnitude difference suggest the components have spectral types of approximately L0.5 and L1.5.

\begin{figure}
 \setlength{\unitlength}{1mm}
 \begin{picture}(100,120)
 \includegraphics{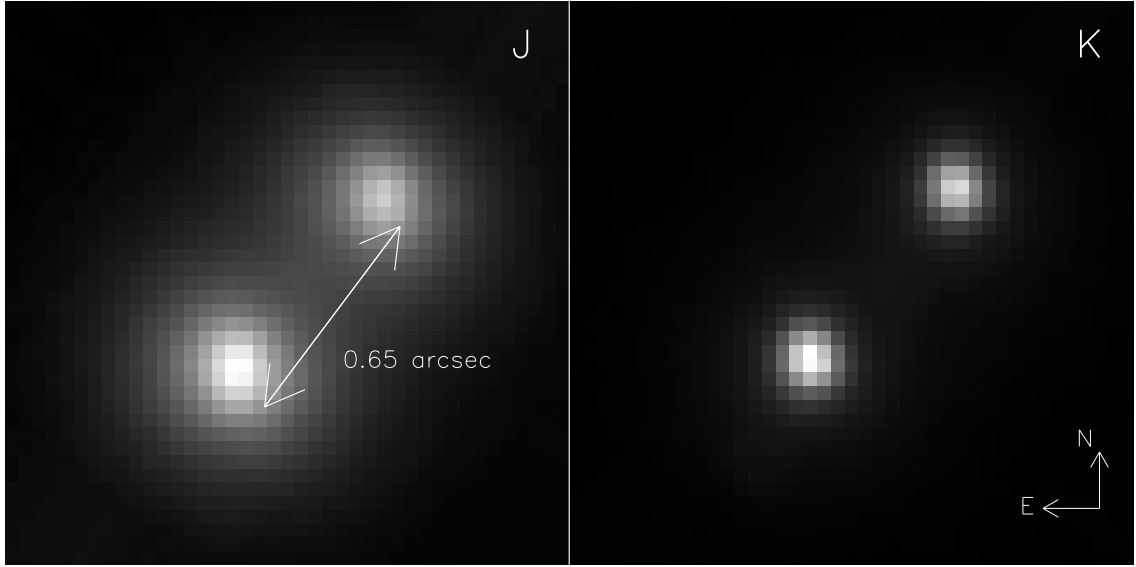}
 \end{picture}
 \caption[]{A Keck/NIRC2 image of 2MASS J05431887+6422528, we find a roughly equal-flux binary with a separation of 0.65\protect\arcsec.} 
  \label{keck0543}
 \end{figure}

\subsubsection{LP 44-334}
\begin{figure}
 \setlength{\unitlength}{1mm}
  \begin{picture}(100,30)
 \includegraphics{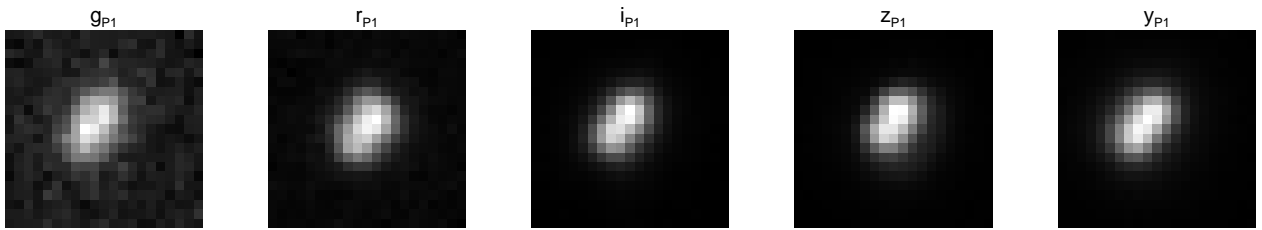}
 \end{picture}
 \caption[]{Pan-STARRS\,1 images of LP 44-334. All images are 5\protect\arcsec\, across with North up and East left. The two components are separated by rough 0.7\protect\arcsec\, with the North-Western component marginally brighter.} 
  \label{LP44_im}
 \end{figure}

We identified the known M6.5 \citep{Reid2004} LP 44-334 as having a strongly positive value of $e_2$ suggesting an elongation in the NW-SE direction (see Figure~\ref{bins6a}). A visual inspection of the best seeing images for each Pan-STARRS\,1 band (see Figure~\ref{LP44_im}) shows the object is almost resolved as a pair elongated in this direction. A visual inspection of the images resulted in a separation estimate of $\sim$0.7\arcsec\, with the NW component appearing to be slightly brighter.

\subsection{Completeness and other detections}
Of the 293 objects where we have enough good data for a fit, we detected every known close binary with a separation greater than 0.3\arcsec\,\, with the exception of DENIS J020529.0$-$115925. We have also examined the individual fits of all the known binaries with separations greater than 0.1\arcsec.\,\. Of these, we find two objects which do not have a large ellipticity at our reference seeing but which show clear trends in ellipticity at very good seeing. 

\begin{figure}
 \setlength{\unitlength}{1mm}
 \begin{tabular}{cc}
 \begin{picture}(40,170)
 \includegraphics{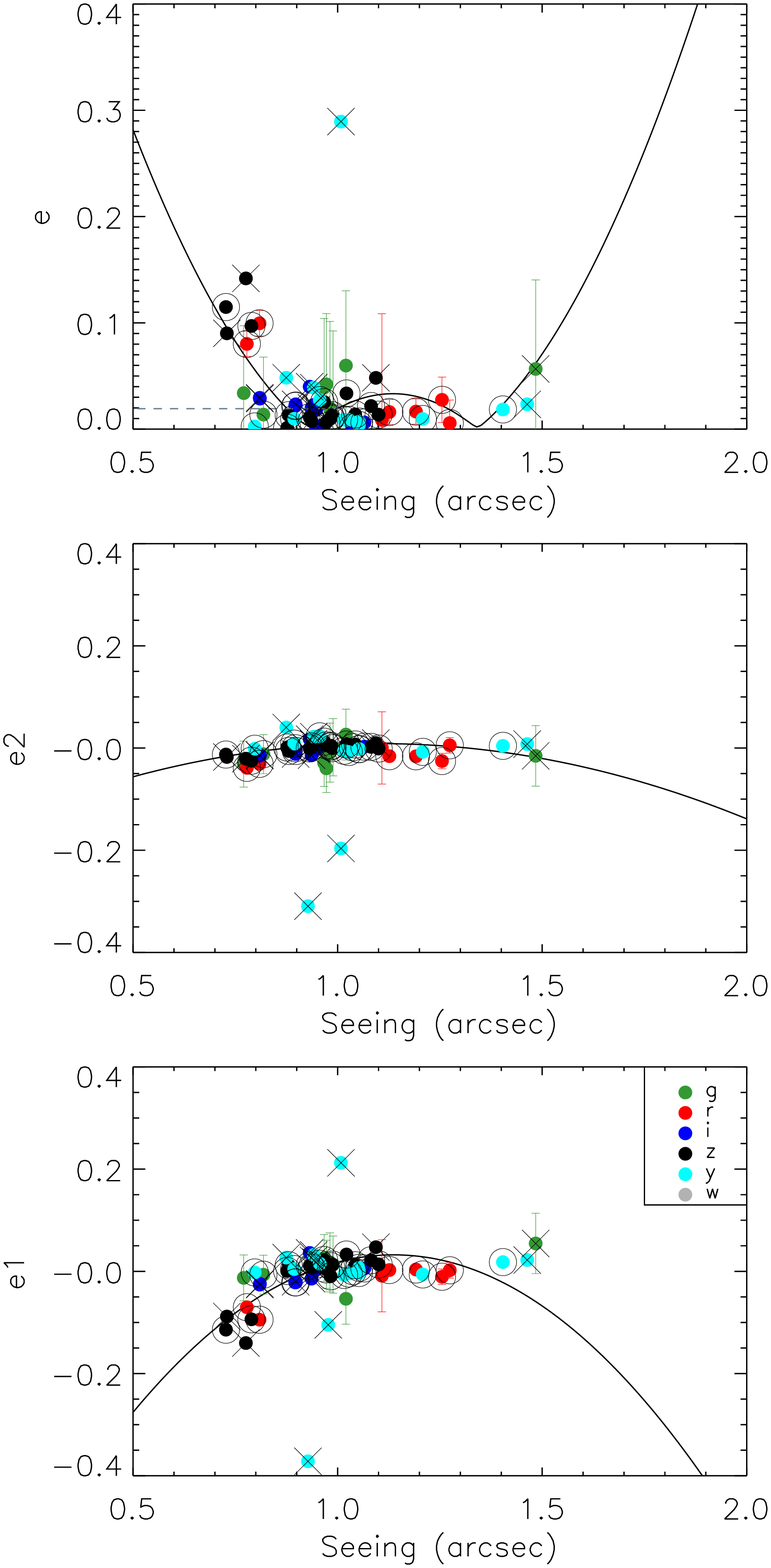}
 \end{picture}&
 \begin{picture}(40,170)
 \includegraphics{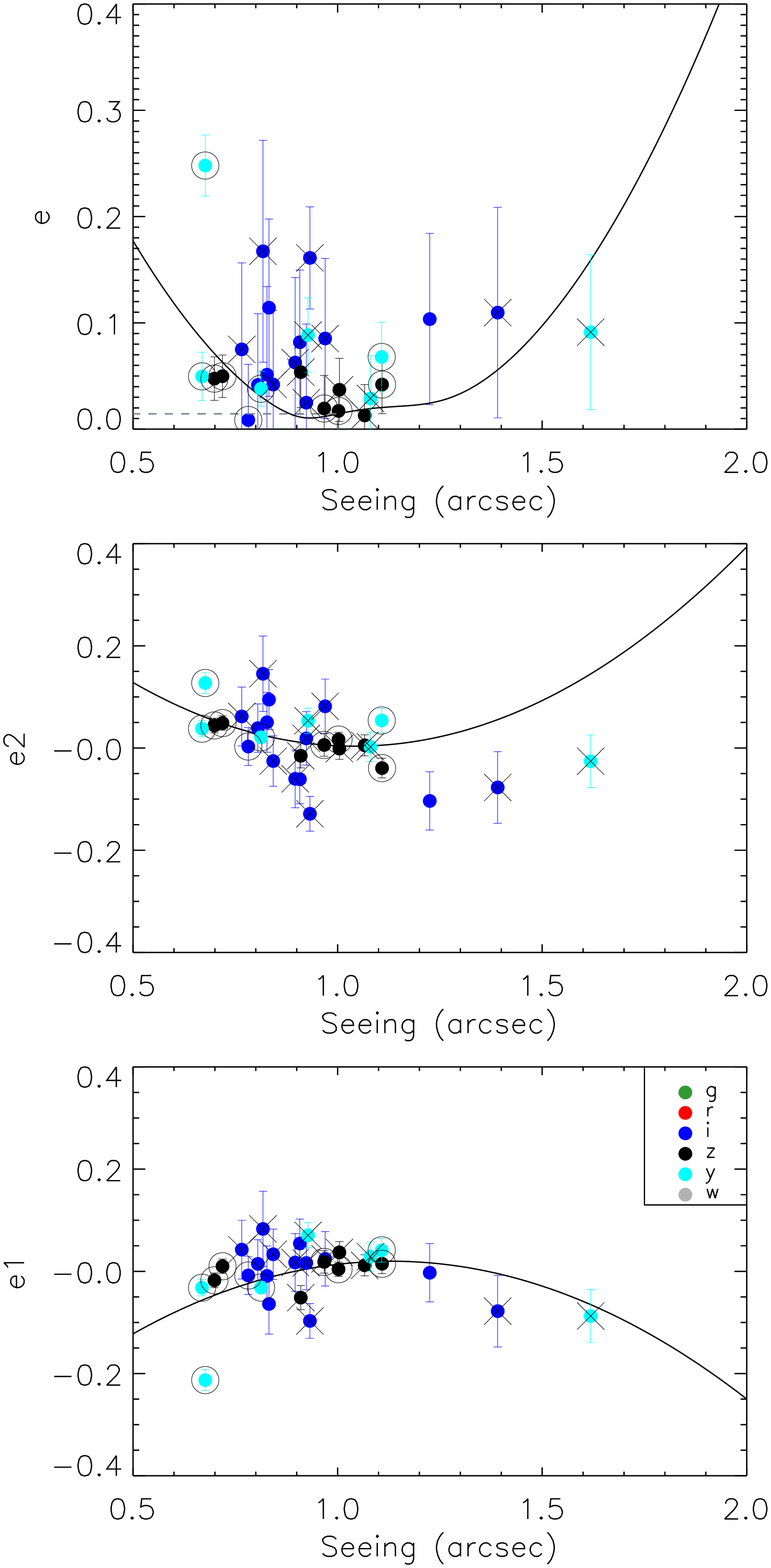}
 \end{picture}
 \end{tabular}
 \caption[]{Ellipticity measurements and fits for the known binaries 2MASSI J0746425+200032 (left) and 2MASS J21522609+0937575 (right). Neither of these objects had a significant ellipticity at the reference seeing but they do show a clear trend at better seeing. The top plot shows the total ellipticity, the middle plot the "x" polarisation $e_2$ and the bottom plot the "+" polarisation $e_2$. Points used in the 2nd order polynomial fits are outlined by a circle and those rejected for data quality reasons are crossed out. Note due to the coordinate system of the postage stamps used a positive $e_2$ is an elongation in the NW-SE direction and a positive $e_1$ is an elongation along the $x$ (i.e. R.A.) axis.} 
  \label{bins7}
 \end{figure}

\subsubsection{2MASSI J0746425+200032}
This L0+L1.5 \cite{Bouy2004} was discovered by \cite{Reid2001b} as a 0.22\arcsec\, binary with a position angle of 15$^{\circ}$. We find (see Figure~\ref{bins7}) that this object has a strongly negative $e_1$ in very good seeing. This suggests an elongation along the Declination axis. Monitoring by M. Liu and T. Dupuy  (Dupuy \& Liu in prep.) suggests a position angle of around 180$^{\circ}$ at a typical Pan-STARRS\,1 epoch of 2012.0, in agreement with our measurement. 
\subsubsection{2MASS J21522609+0937575}
This was detected as an L6+L6 binary by  \cite{Reid2006a} with a separation of 0.25\arcsec\, and a position angle of 106$^{\circ}$. Our fit (see Figure~\ref{bins7}) shows a clear trend in $e_2$ to increasingly positive values whilst $e_1$ shows little trend with the exception of one outlier point. This suggests a position angle closer to 135$^{\circ}$, somewhat different from the measured value in \cite{Reid2006a}. {In their Table~3, \cite{Reid2006a} calculate that 2MASS~J21522609+0937575 is an equal-mass 0.075\,M$_{\odot}$ assuming a 3\,Gyr age. Combining these ages and \cite{Reid2006a}'s separation values and a circular orbit,} we derive an approximate orbital period of 38 years. Whilst we cannot claim to have definitely detected orbital motion, a $\sim$30$^{\circ}$ change in the roughly 5 years between the \cite{Reid2006a} observations and our Pan-STARRS\,1 data is not unreasonable.
\clearpage

\begin{table*}
\begin{minipage}{170mm}
\caption{The individual ellipticity polarisations, implied binary orientation and literature position angles for our sample of recovered binaries.}

\label{known_pa}
\begin{center}
\footnotesize
\begin{tabular}{lccrll}
\hline
Name&$e_{1}$&$e_{2}$&Implied&Literature&Literature\\
&&&Alignment&Separation&P.A.\\
\hline
2MASSI J1426316+155701&$-$0.023$\pm$0.001&0.005$\pm$0.001&N-S&0.32\arcsec&343$^{\circ}$ $^{a}$\\
Kelu-1&$-$0.002$\pm$0.002&$-$0.041$\pm$0.003&NE-SW&0.39\arcsec&255$^{\circ}$ $^{b}$\\
2MASSW J2206228$-$204705&0.042$\pm$0.001&0.006$\pm$0.001&N-S$^*$&0.14\arcsec&$216^{\circ}$ $^{b}$\\
2MASS J09153413+0422045&$-$0.001$\pm$0.005&$-$0.052$\pm$0.006&N-S&0.73\arcsec&$205^{\circ}$ $^{c}$\\
WISEPA J061135.13$-$041024.0&$-$0.044$\pm$0.016&$-$0.046$\pm$0.016&NNE-SSW&0.38\arcsec&33$^{\circ}$ $^{d}$\\
2MASS J17072343$-$0558249&$-$0.007$\pm$0.001&$-$0.065$\pm$0.001&N-S&1.00\arcsec&35$^{\circ}$ $^{c}$\\
SIMP J1619275+031350&0.062$\pm$0.025&$-$0.108$\pm$0.025&ENE-WSW&0.69\arcsec&71$^{\circ}$ $^{e}$\\
DENIS-P J220002.05$-$303832.9&$-$0.211$\pm$0.008&$-$0.036$\pm$0.009&N-S&1.09\arcsec&177$^{\circ}$ $^{f}$\\
2MASS J15500845+1455180&$-$0.229$\pm$0.011&$-$0.030$\pm$0.011&N-S&0.91\arcsec&17$^{\circ}$ $^{g}$\\
DENIS J020529.0$-$115925&0.785$\pm$0.161&$-$0.004$\pm$0.161&E-W&0.29\arcsec&246$^{\circ}$ $^{h\dag}$\\
\hline
\multicolumn{6}{l}{$^{a}$ \protect\cite{Konopacky2010}} \\
\multicolumn{6}{l}{$^{b}$ Dupuy \& Liu orbital monitoring  (Dupuy \& Liu in prep.)} \\
\multicolumn{6}{l}{$^{c}$ \protect\cite{Reid2006a}} \\
\multicolumn{6}{l}{$^{d}$ \protect\cite{Gelino2014}} \\
\multicolumn{6}{l}{$^{e}$ \protect\cite{Artigau2011}} \\
\multicolumn{6}{l}{$^{f}$ \protect\cite{Burgasser2006a}} \\
\multicolumn{6}{l}{$^{g}$ \protect\cite{Burgasser2009a}} \\
\multicolumn{6}{l}{$^{h}$ \protect\cite{Bouy2003}} \\
\multicolumn{6}{l}{$^{*}$ Whilst this object has reference seeing ellipticities which suggest an E-W alignment, inspection of the actual fit (Figure \protect\ref{bins2}) } \\
\multicolumn{6}{l}{$^{ }$shows four points with excellent seeing with $Ð$ve $e_2$ values, suggesting N-S.}\\
\multicolumn{6}{l}{$^{\dag}$ Spurious ellipticity estimate due to exceptionally good seeing on a handful of measurements.} \\

\normalsize
\end{tabular}
\end{center}
\end{minipage}
\end{table*}

\section{Application to the full Pan-STARRS\,1 }
\label{db_test}
We applied the methods set out above to the Pan-STARRS\,1 database. The database table {\verb FORCED_WARP_LENS } includes the relevant parameters to estimate the the shapes of objects and to correct for PSF anisotropy. The image moments $M_{XX}$, $M_{XY}$ and $M_{YY}$ are equivalent to the parameters $I_{11}$, $I_{12}$ and $I_{22}$  respectively. The PSF anisotropy was measured by dividing each skycell in the image into 5'$\times$5' areas. The median image parameters of PSF stars in each region were then determined and recorded in the database with the suffix {\verb _PSF }. Thus once one has calculated $e_1$ and $e_2$ from the image moments the appropriate anisotropy correction can be made using,

\begin{equation}
\begin{array}{ll}
P_{\nu,\nu}&=X\nu\nu\_sm\_OBJ-e\nu\_SM\_OBJ \times e_{\nu}\\
P_{smooth, \nu,\nu}&=X\nu\nu\_sm\_PSF-e\nu\_SM_PSF \times e\nu\_PSF\\

p_{smooth,\nu}&=\frac{e\nu\_PSF}{P_{smooth,\nu,\nu}}\\
e_{\nu,corr}&=e_{\nu}-p_{smooth,\nu}\times P_{\nu,\nu}
\end{array}
\end{equation}

The value of $\nu$ here can either be 1 or 2 so for example the final equation above is equivalent to Equation~\ref{e_cor}. Parameters such as $X\nu\nu\_sm\_OBJ$ are a shorthand for the database parameters $X11\_sm\_OBJ$ or $X22\_sm\_OBJ$, etc.

\begin{figure}
 \setlength{\unitlength}{1mm}
 \begin{picture}(100,100)
 \includegraphics{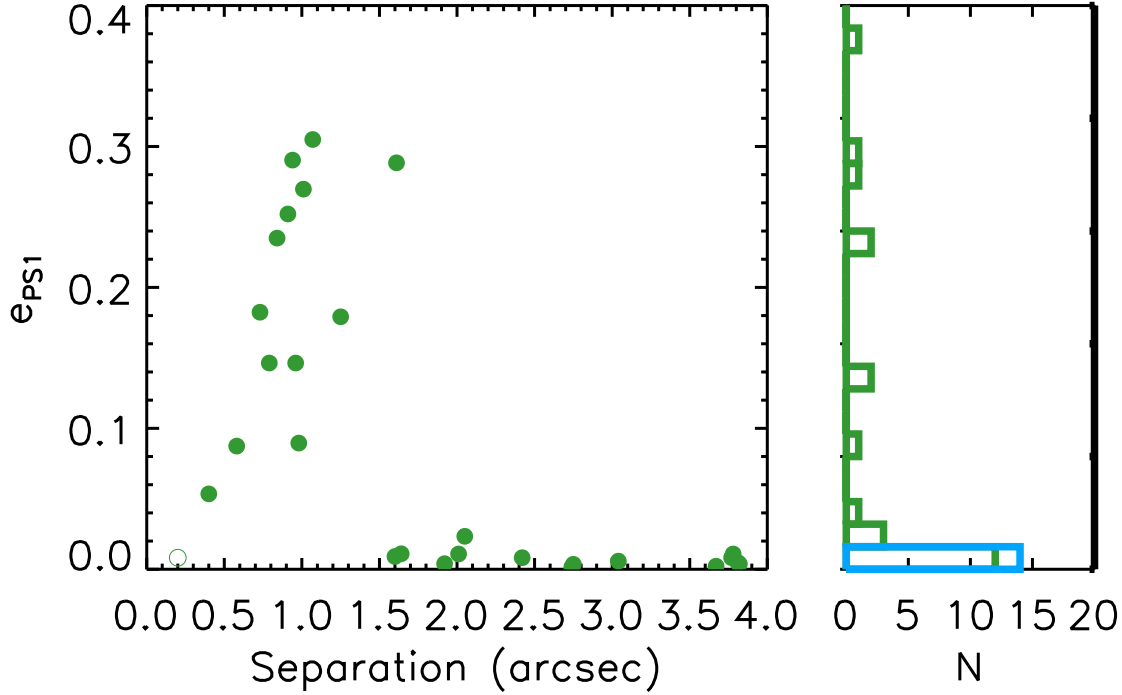}
 \end{picture}
 \caption[]{Left: the measured ellipticites at our reference seeing of 1\arcsec\, for binaries in the test sample of \protect\cite{Terziev2013} produced using the image shape parameters from the Pan-STARRS\,1 database alone. The open circle is an object with a very small separation which Terziev notes would not have affected the ellipticity measurement. Right: a histogram of these ellipticities for binaries (green) and single stars (blue). Note the one binary at a separation of 2\arcsec\, which has a low elliipticity. This has the smallest flux ratio in the sample at 0.04.} 
  \label{terziev_pv3_test}
 \end{figure}

We extracted the detections for the sample of stars from \cite{Terziev2013}. We then applied the correction procedure to each image to produce PSF anisotropy corrected shape measurements for each object. The results are shown in Figure~\ref{terziev_pv3_test}. Clearly we are able to detect the closer ($<$1.5\arcsec) binaries but do not measure any significantly distorted images from the wider binaries. The reason for this is that the Pan-STARRS\,1 database shape measurement database calculations are based on the position of the primary star in the stacked Pan-STARRS\,1 images. Conversely in our image-based method we used positions from SExtractor with no deblending, in reality intentionally blurring the image so the centroid position of the blended image is used as the central position for shape measurement calculations. Recall that this was to stop the centroid used for shape measurement changing if the secondary was resolved by SExtractor in a handful of images. This would result in an image centroid between the two components thus producing higher image ellipticities than the case where the centroid is on the centre of the brighter binary component as the flux from the secondary will be more strongly suppressed by the weighting function. Note that all of the test sample binaries wider than 2.01\arcsec\, had their secondary component detected as an individual star in the Pan-STARRS\,1 database. However there is a blindspot for this method from the current database between 1.5\arcsec\, and 2\arcsec\, where we would miss a substantial number of binaries.

\subsection{Correction for centroiding errors}
Our Pan-STARRS\,1 database shape measurements rely on the positions of the objects in the stacked image data. For objects with little or no proper motion this will produce accurate ellipticity measurements. However if the object moves over the course of the Pan-STARRS\,1 survey then the ellipticity measurement will be artificially raised. This is because the image moments will be calculated relative to a position which will not represent the centroid of each individual detection. This implies that the error in the ellipiticity should rise as $\Delta pos^2$. There is however another factor, the weighting function which excludes flux far away from the centroid used in the calculation. 

First we must consider how the positional offset affects the two ellipticity polarisations,   
\begin{equation}
\begin{array}{ll}
\Delta pos_1& = - \mu \Delta t  \cos \left( \frac{\pi P.A.}{90} \right)\\
\Delta pos_2& = - \mu \Delta t  \sin \left( \frac{\pi P.A.}{90} \right)\\
\end{array}
\end{equation}
We simulated the offset in ellipticity caused by positional offset by taking a Gaussian PSF and moving the centroid position about which we measure the flux distribution. We used a series of different of seeing $FWHM$ values and found that the change in ellipticity was well modelled by,

\begin{equation}
|\Delta e| = \frac{(\Delta pos/FWHM)^2}{0.08+(\Delta pos/FWHM)^2}
\end{equation}
Thus
\begin{equation}
\begin{array}{l}
\Delta e_1 = \sgn(\Delta pos_1) \frac{(\Delta pos_1/FWHM)^2}{0.08+(\Delta pos_1/FWHM)^2}\\
\Delta e_2 = \sgn(\Delta pos_2) \frac{(\Delta pos_2/FWHM)^2}{0.08+(\Delta pos_2/FWHM)^2}\\
\end{array}
\end{equation}
We note that these are approximations and are likely only useful for small offsets. Our work on the Terziev test sample (which typically have proper motions below 0.1\arcsec/yr)\, shows that binaries can be reliably detected for low proper motion objects without corrections. We would strongly caution that shape measurements for higher proper motion stars may be unreliable even after applying a correction factor. 
\section{Conclusions}
We have applied shape measurement techniques to recover previously known binaries and discover three new binaries. We show that this method can reliably recover binaries wider than around 0.3\arcsec.\, The Pan-STARRS\,1 database includes an implementation of stellar shape measurements which will hopefully become available in a future data release. These data will allow efficient screening of adaptive optics observations for close binaries. Future large surveys such as LSST would benefit from the availability of individual observation anisotropy parameters from all stellar objects. This would allow the study of large samples of partially resolved binaries to probe stellar multiplicity.
\section*{Acknowledgments}
The Pan-STARRS1 Surveys (PS1) have been made possible through contributions of the Institute for Astronomy, the University of Hawaii, the Pan-STARRS Project Office, the Max-Planck Society and its participating institutes, the Max Planck Institute for Astronomy, Heidelberg and the Max Planck Institute for Extraterrestrial Physics, Garching, The Johns Hopkins University, Durham University, the University of Edinburgh, Queen's University Belfast, the Harvard-Smithsonian Center for Astrophysics, the Las Cumbres Observatory Global Telescope Network Incorporated, the National Central University of Taiwan, the Space Telescope Science Institute, the National Aeronautics and Space Administration under Grant No. NNX08AR22G issued through the Planetary Science Division of the NASA Science Mission Directorate, the National Science Foundation under Grant No. AST-1238877, the University of Maryland, and Eotvos Lorand University (ELTE). Partial support for this work was provided by National Science Foundation grants AST-1313455 (EAM \& WMJB) and NSF-AST-1518339 (MCL \& WMJB). The authors would like to thank Adam Kraus, Emil Terziev, Nick Law, Thomas Dixon and Nick Kaiser for useful discussions.

\bibliography{../ndeacon}

\bibliographystyle{mn2e}
\label{lastpage}

\end{document}